\documentclass[english,amsmath,amssymb,superscriptaddress,twocolumn,prb]{revtex4}
\usepackage{graphicx}
\usepackage{dcolumn}
\usepackage{bm}

\begin{document}

\title{
Equation-of-Motion Series Expansion of Double-Time Green's Functions }

\author{Ning-Hua Tong}
\affiliation{ Department of Physics, Renmin University of China, 100872 Beijing, China }
\affiliation{ Beijing Key Laboratory of Opto-electronic Functional Materials and Micro-nano Devices (Renmin University of China)}
\date{\today}

\begin{abstract}
Based on the Green's function (GF) equation-of-motion formalism, we develop a method to expand the double time Green's function into Taylor series of the parameter $\lambda$ in the Hamiltonian $H=H_0 + \lambda H_1$. Here $H_0$ is the exactly solvable part and $H_1$ is regarded as the perturbation. To restore the analytical structure of GF, we use the continued fraction to do resummation for the obtained series. The problem of zero-temperature divergence is identified and remedied by the self-consistent series expansion. To demonstrate the implementation of this method, we carry out the weak- as well as the strong-coupling expansion for the Anderson impurity model to order  $\lambda^2$. Improved result for the local density of states is obtained by self-consistent second-order strong-coupling expansion and continued fraction resummation. 
\end{abstract}


\maketitle
\section{Introduction}

Green's function (GF) is widely used in the study of quantum many-body problems in condensed matter physics.\cite{Abrikosov1} It is not only a common language to describe the fundamental physical concepts and processes, but also an important tool to do quantitative calculations for physical observables. Among all the methods of calculating a GF, expanding it into the power series of certain parameter $\lambda$ is a basic and straightforward method for the Hamiltonian $H = H_0 + \lambda H_1$, provided that $H_0$ is exactly solvable and its GF obtainable. In cases where no other reliable results are available for GF, such a series expansion (SE) provides a reference which is accurate in the limit of small $\lambda$. Besides quantitative information, important qualitative understanding of the system can also be obtained by analysing the properties of GF series. Well-known examples are the Fermi liquid properties of weakly interacting fermions\cite{Luttinger1} and the simplification of theory for lattice fermions in the large spatial dimension limit.\cite{Mueller1,Metzner1}

Various GF series expansion methods have been developed so far. For a weakly interacting system, the weak-coupling expansion applies, where the non-interacting part of the Hamiltonian is chosen as $H_0$ and $H_1$ is the interaction part. Using Wick's theorem, interacting GF is expanded in terms of the interacting vertex and free GF. Standard Feynman diagram techniques facilitate the representation and calculation of the series. Various partial summation methods have been developed diagrammatically, including Hartree-Fock, random phase approximation, and fluctuation-exchange approximation.\cite{Mattuck1}

In the other limit where the interaction is strong, the strong-coupling expansion can be considered, if the interacting part of Hamiltonian is exactly solvable. In this case, $H_0$ and $H_1$ denote the interacting part and the non-interacting part of a given Hamiltonian, respectively. Although the conventional Wick's theorem no longer holds, various diagram techniques have been developed. Wick-like theorems were established\cite{Slobodyan1,Yang1,Shvaika1} for the standard basis operators\cite{Haley1} (an extension of the Hubbard operators\cite{Hubbard1}) to develop diagrammatic expansions for GF. A more convenient way is the cumulant expansion method\cite{Kubo1} introduced by Metzner. He expands the GF in terms of the hopping lines and local cumulants\cite{Metzner2} with unrestricted summations. In another elegant approach, the strong-coupling problem of original fermions is transformed into an effective weak-coupling problem of the dual fermions through a Grassmann Hubbard-Stratonovich transformation.\cite{Sarker1} The GF of the dual fermions and of the original fermions are then obtained from standard Feynmann diagram technique.\cite{Pairault1,Pairault2,Brune1,Dupuis1} In both the weak- and the strong-coupling expansion approaches, it is necessary to sum partial contributions in the series to infinite order with a resummation method to restore the analytical structure of GF.\cite{Pairault1,Sherman1} If done correctly, such resummation can significantly extend the validity range of the theory.

In recent years, Monte Carlo (MC) sampling methods have been used to carry out the series expansion of GFs, either in the form of determinant calculation\cite{Gull1,Iazzi1} or the direct diagram summations.\cite{Houcke1} In these methods, usually a large number of expansion terms can be sampled and summed to give the GF which is reliable in broad parameter regimes. 

In this work, we develop a method for expanding the double time GF into power series of a given parameter $\lambda$, based on its equations of motion (EOM). We call this method EOM series expansion. Compared to the methods summarized above, this approach is distinctive due to following features. First, it is universal in the sense that the formalism does not depend on the concrete form of $H_0$. For any Hamiltonian $H_0$ whose eigen states and eigen energies, or GF can be obtained exactly, one can always expand the full GF in terms of $H_1$ in a recursive fashion. Second, there is no restriction in the operators $A$ and $B$ that define the double time GF $G(A|B)_{\omega}$. Single particle as well as many-particle GFs with two time variables can be obtained in the same general formalism. Third, the expansion calculation only involves double time GFs of $H_0$. The complicated multiple-time integrations in traditional expansion method are replaced with operator commutator calculations here. Fourth, in the present framework, the issue of partial infinite summation and the zero temperature divergence problem in the unrenormalized series expansion can be dealt with by a standard procedure, i.e., by using the continued fraction resummation and the self-consistent expansion scheme. Fifth, EOM of the residue of a finite order expansion is given, providing a possible means to estimate the error of the truncated series and to improve the expansion result.

In this paper, we carry out the EOM expansion of single particle GF for the Anderson impurity model. Both the weak-coupling expansion to $U^2$ order and the strong-coupling expansion to $V_{k}^{2}$ order are obtained for single particle GF. In this work, we put the emphasis on the strong-coupling expansion. By comparing the obtained local density of states (LDOS) with that from the numerical renormalization group (NRG) method, we evaluate the effect of the bare EOM expansion and the self-consistent EOM expansion, supplemented with different resummation methods. We show that the self-consistent EOM expansion together with the continued fraction resummation gives qualitatively correct results which are improved in several aspects over previous ones.

This paper is organized as follows. In Sec. II, we present the formal formalisms, including the the double time GF EOM series expansion, the resummation methods using self-energy and continued fraction, and the self-consistent EOM expansion. In Sec. III, the single impurity Anderson model is studied by this method. The weak- and strong-coupling expansions are carried out to second order, respectively. The strong-coupling expansion results obtained from different resummation methods are compared with NRG results. In Sec. IV, several issues about the method are discussed and a summary is given.

\section{Equations-of-Motion Series Expansion of Double Time Green's Functions}

\subsection{Equations of Motion Series Expansion}
We start from the EOM of retarded GFs. Let us consider the following retarded GF defined by two operators $A$ and $B$ at two times $t$ and $t^{\prime}$, respectively,
\begin{equation}  \label{eq:1}
  G^{r}\left[A(t)|B(t^{\prime}) \right] \equiv \frac{1}{i} \theta(t-t^{\prime}) \langle \left[A(t), B(t^{\prime}) \right]_{\pm} \rangle
\end{equation}
Here, $\theta(x)$ is the step function. $O(t) = e^{iHt}Oe^{-iHt}$ is the Heisenberg operator with respect to the Hamiltonian $H$. $\left[X, Y \right]_{\pm} = XY \pm YX$. The plus sign is for fermion-type GF, and the minus sign for boson-type GF, respectively. $\langle O \rangle = Tr \left(e^{-\beta H}O \right)/Tre^{-\beta H}$ is the average in thermal equilibrium state of $H$. Here $\hbar=k_{B}= 1$ is used. In this paper, the target of expansion is the GF defined in Eq.(\ref{eq:1}) with only two time variables. We focus on the equilibrium state where the GF depends only on $t-t^{\prime}$, although the method can be generalized to the non-equilibrium case. In the equilibrium state, the Fourier transformation of $G^{r}\left[A(t)|B(t^{\prime}) \right]$ will be denoted as $G(A|B)_{\omega+ i\eta}$,
\begin{equation}   \label{eq:2}
  G(A|B)_{\omega + i \eta}= \displaystyle \int_{-\infty}^{\infty}   G^{r}\left[A(t)|B(t^{\prime}) \right] e^{i (\omega + i\eta)(t-t^{\prime}) } d(t-t^{\prime}).
\end{equation}
Here $\eta$ is an infinitesimal positive number to guarantee the convergence of integration.
 
Calculating the derivative of Eq.(\ref{eq:1}) with respect to $t$ or $t^{\prime}$ and transforming it onto frequency axis, one easily obtains the EOM for the double time GF as
\begin{eqnarray}   \label{eq:3}
  \omega G(A|B)_{\omega} &=& \langle \left[A, B \right]_{\pm} \rangle + G([A, H]|B)_{\omega} \nonumber \\
     &=&\langle \left[A, B \right]_{\pm} \rangle - G(A| [B, H] )_{\omega}.
\end{eqnarray}
On the right hand side of Eq.(\ref{eq:3}), new operators emerge from the commutator $[A,H]$ or $[B, H]$ and the GFs defined by them usually involve more particles. When the EOM for these new GFs are written down, even higher order operators and corresponding GFs will be generated. Usually this heirarchical EOM can not close automatically and approximate truncations have to be introduced to form a closed set of algebraic equations. In this way, GFs will be expressed explicitly in terms of $\omega$ and some unknown averages. Finally, these averages will be calculated self-consistently from GFs through the fluctuation-dissipation theorem,
\begin{equation}   \label{eq:4}
  \langle BA \rangle = -\frac{1}{\pi} \displaystyle\int_{-\infty}^{\infty} {\text Im} G(A|B)_{\omega+ i\eta} \frac{1}{e^{\beta \omega} \pm 1} d\omega.
\end{equation}

Being flexible and non-perturbative, the above EOM formalism has been widely used in the study of quantum many-body systems since the early works of Bogoliubov,\cite{Bogolyubov1} Anderson,\cite{Anderson1} Hubbard\cite{Hubbard2} and others. The applications range from Kondo physics\cite{Luo1} to quantum magnetism.\cite{Rudoi1,Callen1} However, due to the lack of a universal and systematic truncation scheme, it is difficult to control the precision of the resulting GFs. Especially, the analytical structure of GF may be violated by the truncation. Usually, well established truncation schemes are obtained empirically and only apply to specific problems. Here, we will employ the EOM formalism to obtain a systematic series expansion for the double time GFs.

We first discuss the type of Hamiltonian which is exactly solvable in the context of EOM. For a large class of Hamiltonians and operators $A$ and $B$, the hierarchy of EOM Eq.(\ref{eq:3}) can form a closed set of algebraic equations. The GFs appearing in this set can be solved exactly even in the thermodynamical limit. Such Hamiltonians include, for examples, the non-interacting Hamiltonian of free electrons on a lattice $H_0 = \sum_k \epsilon_k c_{k}^{\dagger}c_{k}$, and the Hubbard model in the atomic limit $H_0 = U \sum_i n_{i \uparrow} n_{i \downarrow}$. The hierarchical EOM naturally close for these Hamitonians. For such exactly solvable Hamiltonians, the corresponding super-operator $\hat{\hat{\mathcal{L}}}$, defined as $\hat{\hat{\mathcal{L}}} O \equiv \left[H_0, O \right]$, has certain symmetries and hence has finite dimensional invariant subspaces even in the thermodynamical limit. $G_{0}(A|B)_{\omega}$ can be solved exactly if operators $A$ and $B$ belong to the subspace.

In general cases where the closure of EOM is not obvious, EOM of GFs can still be solved exactly if every eigen-state $|\mu \rangle$ and eigen energy $E_{\mu}$ of $H_0$ can be obtained, $H_0 |\mu \rangle = E_\mu |\mu \rangle$. In such cases one can construct the standard basis operators (SBOs)\cite{Haley1} $A_{\alpha \beta} \equiv | \alpha \rangle \langle \beta|$. Any operator in the Hilbert space of $H_0$ can be expanded by SBOs as
\begin{eqnarray}    \label{eq:5}
   && A = \displaystyle\sum_{\alpha \beta} f_{\alpha \beta}A_{\alpha \beta}, \nonumber \\
   && B = \displaystyle\sum_{\alpha \beta} g_{\alpha \beta}A_{\alpha \beta}.
\end{eqnarray}
Here $f_{\alpha\beta} = \langle \alpha|A|\beta\rangle$ and $g_{\alpha\beta} = \langle \alpha|B|\beta\rangle$. The EOM of GFs defined by SBOs $G_{0}(A_{\alpha \beta}|A_{\mu \nu})_{\omega}$ naturally close and give
\begin{equation}   \label{eq:6}
    G_{0}(A_{\alpha \beta}|A_{\mu \nu})_{\omega} = \delta_{\beta\mu} \delta_{\alpha \nu} \frac{ \langle A_{\alpha \alpha}\rangle \pm  \langle A_{\beta \beta}\rangle}{\omega + E_{\alpha} - E_{\beta}},
\end{equation}
where $\langle A_{\alpha \alpha }\rangle = e^{-\beta E_{\alpha}}/Z_0$ and $Z_0$ is the partition function of $H_0$. $G_{0}(A|B)_{\omega}$ is then obtained as $G_{0}(A|B)_{\omega} = \sum_{\alpha \beta} \sum_{\mu \nu} f_{\alpha\beta} g_{\mu\nu} G(A_{\alpha \beta}|A_{\mu \nu})_{\omega}$.

Now suppose that the full Hamiltonian is a sum of an exactly solvable part $H_0$ and a perturbation $H_1$. We add a bookmarking factor $\lambda$ to $H_1$ and the Hamiltonian reads $H= H_0 + \lambda H_1$. We will expand $G(A|B)_{\omega}$ into a power series of $\lambda$ and set $\lambda$ as unity afterwards. Formally, the Taylor series expansion of GF and averages read
\begin{eqnarray}   \label{eq:7}
&& G(A|B)_{\omega}  \nonumber \\
& =& G_0(A|B)_{\omega} + \lambda G_1(A|B)_{\omega} + ... + \lambda^n G_n(A|B)_{\omega}  \nonumber \\
  && + \Gamma_n(A|B)_{\omega},
\end{eqnarray}
and 
\begin{equation}   \label{eq:8}
  \langle O \rangle = \langle O \rangle_0 + \lambda \langle O \rangle_1 + ... + \lambda^{n} \langle O \rangle_n + \langle O \rangle^{R}_n.
\end{equation}
Here $G_i(A|B)_{\omega}$ and $\langle O \rangle_i$ are the $(H_1)^{i}$-order contributions to GF and average, respectively. $\Gamma_n(A|B)_{\omega} \sim O(\lambda^{n+1})$ and $\langle O \rangle^{R}_{n} \sim O(\lambda^{n+1})$ are the residues of this expansion up to order $n$. 

Expanding the GFs and averages in Eq.(\ref{eq:3}) and comparing the coefficients of $\lambda^{i}$ on both sides of equations, one gets for $i \geq 1$
\begin{eqnarray}   \label{eq:9}
&& \omega G_i(A|B)_{\omega}  \nonumber \\
&& = \langle \left[A, B \right]_{\pm} \rangle_i + G_{i-1}([A, H_1]|B)_{\omega} + G_{i}([A, H_0]|B)_{\omega} \nonumber \\
&& = \langle \left[A, B \right]_{\pm} \rangle_i - G_{i-1}(A|[B, H_1])_{\omega} - G_{i}(A|[B, H_0])_{\omega}, \nonumber \\
&&
\end{eqnarray}
and for $i=0$ 
\begin{eqnarray}   \label{eq:10}
 \omega G_0(A|B)_{\omega} &=& \langle \left[A, B \right]_{\pm} \rangle_0 + G_0([A, H_0]|B)_{\omega}   \nonumber \\
&=& \langle \left[A, B \right]_{\pm} \rangle_0 - G_0(A|[B, H_0])_{\omega}  .
\end{eqnarray}
The residue $\Gamma_n(A|B)_{\omega}$ of the $n$-th order expansion satisfies the EOM 
\begin{eqnarray}   \label{eq:11}
&& \omega \Gamma_n(A|B)_{\omega}  \nonumber \\
&& = \langle \left[A, B \right]_{\pm} \rangle^{R}_n + G_{n}([A, H_1]|B)_{\omega} + \Gamma_{n}([A, H]|B)_{\omega}. \nonumber \\
&& = \langle \left[A, B \right]_{\pm} \rangle^{R}_n - G_{n}(A|[B, H_1])_{\omega} - \Gamma_{n}(A|[B, H])_{\omega}. \nonumber \\
&&
\end{eqnarray}
In this formalism, one can choose to use the left-side or the right-side EOM formula differently for different order $i$.

To solve the averages involved in the above equations, expanding Eq.(\ref{eq:4}) gives
\begin{equation}   \label{eq:12}
  \langle BA \rangle_i = -\frac{1}{\pi} \displaystyle\int_{-\infty}^{\infty} {\text Im} G_i(A|B)_{\omega+ i\eta} \frac{1}{e^{\beta \omega} \pm 1} d\omega
\end{equation}
for $0 \leq i \leq n$. The residue $\langle O \rangle^R_n$ is obtained from
\begin{equation}   \label{eq:13}
  \langle BA \rangle^{R}_n = -\frac{1}{\pi} \displaystyle\int_{-\infty}^{\infty} {\text Im} \Gamma_n(A|B)_{\omega+ i\eta} \frac{1}{e^{\beta \omega} \pm 1} d\omega.
\end{equation}

Since $H_0$ is exactly solvable in the sense discussed above, the EOM of $G_{i}(A|B)_{\omega}$ in Eq.(\ref{eq:9}) will close because the series $A$, $[A, H_0]$, $[[A, H_0], H_0]$, ... generates closed set of operators. In cases where this is not obvious or too complicated, one could decompose $A$ and $B$ into SBOs and study the Taylor series expansion for the GFs defined with SBOs. In any case, $G_{i}(A|B)_{\omega}$ can be expressed in terms of the lower order GF $G_{i-1}(A^{\prime}|B)_{\omega}$ with more complicated operators $A^{\prime}$. Repeatedly employing Eq.(\ref{eq:9}), one can then reduce $G_{i-1}(A^{\prime}|B)_{\omega}$ to $G_{i-2}(A^{\prime \prime}|B)_{\omega}$, and so on. Finally the GF component of every order $i \geq 1$ can be reduced to the type $G_0(A|B)$ with different operators $A$. These zeroth order GFs are exactly solvable. Therefore, Eq.(\ref{eq:9}) provides a practical way of calculating arbitrary order contributions to $G(A|B)_{\omega}$. 

The $n$-th order residue of the expansion $\Gamma_n(A|B)_{\omega}$ is determined by its EOM Eq.(\ref{eq:11}) which cannot be solved exactly. By truncating the hierarchical EOM Eq.(\ref{eq:11}), one could obtain an approximate result for $\Gamma_n(A|B)_{\omega}$. This result could be used to evaluate or accelerate the convergence of the expansion.

Equation (9) contains $\langle [A, B]_{\pm}\rangle_{i}$, the $i$-th order contribution to $\langle [A, B]_{\pm}\rangle$. It must be calculated through Eq.(\ref{eq:12}), which leads to a set of algebraic equations for the averages. In the conventional GF EOM approach with truncation approximations, neither consistency nor sufficiency is guaranteed for this set of equations. In the rigorous series expansion presented here, for each order $i$, the set of equations is both consistent and sufficient. That is, a unique solution of the averages at $i$-th order is always obtainable. The $n$-th order residue of the average $\langle [A, B]_{\pm} \rangle^{R}_n $ needs to be calculated self-consistently using Eq.(\ref{eq:13}).

\subsection{Resummation Methods}

Truncating the Taylor series of a GF at a finite order always produces the problem of causality. This is most easily demonstrated by Taylor expanding the Lehmann representation of $G(A|B)_{\omega}$ (taking the fermion-type GF as an example)
\begin{equation}    \label{eq:14}
    G(A|B)_{\omega} = \frac{1}{Z} \displaystyle\sum_{m,n} \frac{e^{-\beta E_m} + e^{-\beta E_n}}{\omega + E_m - E_n} X_{mn}.
\end{equation}
Here $|m\rangle$ and $E_m$ are the eigen state and the eigen energy of $H$, respectively. $Z = \sum_{m}e^{-\beta E_m}$ is the partition function, and $X_{mn} = \langle m | A | n \rangle \langle n | B | m \rangle$ is the matrix element. It is seen that GF has only real simple poles and $\beta$ appears on the exponent.

We can formally expand $Z$, $E_m$ and $X_{mn}$ into power series of $\lambda$ to  obtain 
\begin{eqnarray}   \label{eq:15}
 && Z = \sum_{i=0}^{\infty} \lambda^{i} Z_{i} \nonumber \\
 && E_m = \sum_{i=0}^{\infty} \lambda^{i} E_{m}^{(i)} \nonumber \\
 && X_{mn} = \sum_{i=0}^{\infty} \lambda^{i} X_{mn}^{(i)}.
\end{eqnarray}
The Lehmann representation for the $i$-th order term $G_{i}(A|B)_{\omega}$ can be obtained by putting these expansions into Eq.(\ref{eq:14}) and collecting terms proportional to $\lambda^{i}$. The first two orders read
\begin{equation}   \label{eq:16} 
    G_{0}(A|B)_{\omega} = \frac{1}{\displaystyle \sum_{m}e^{-\beta E^{(0)}_m}} \displaystyle\sum_{m,n} \frac{e^{-\beta E^{(0)}_m} + e^{-\beta E^{(0)}_n}}{\omega + E^{(0)}_m - E^{(0)}_n} X^{(0)}_{mn},
\end{equation}
and 
\begin{eqnarray}   \label{eq:17}
&&    G_{1}(A|B)_{\omega}  \nonumber \\
&& = \frac{1}{Z_0} \displaystyle\sum_{m,n} \frac{e^{-\beta E^{(0)}_n} \left[ -\frac{Z_1}{Z_0} X_{mn}^{(0)} - \beta E_{n}^{(1)}X_{mn}^{(0)} + X_{mn}^{(1)} \right]}{\omega + E^{(0)}_m - E^{(0)}_n}   \nonumber \\ 
&& +     \frac{1}{Z_0} \displaystyle\sum_{m,n} \frac{e^{-\beta E^{(0)}_m} \left[ -\frac{Z_1}{Z_0} X_{mn}^{(0)} - \beta E_{m}^{(1)}X_{mn}^{(0)} + X_{mn}^{(1)} \right]}{\omega + E^{(0)}_m - E^{(0)}_n}   \nonumber \\
&& -   \frac{1}{Z_0} \displaystyle\sum_{m,n} \frac{ \left[ e^{-\beta E^{(0)}_m} + e^{-\beta E^{(0)}_n} \right] \left[ E_{m}^{(1)} - E_{n}^{(1)} \right]}{ \left[\omega + E^{(0)}_m - E^{(0)}_n \right]^2 } X_{mn}^{(0)}.  \nonumber \\
&&
\end{eqnarray}

Eq.(\ref{eq:16}) shows that $G_{0}(A|B)_{\omega}$ has real simple poles which gives the expected causality of the retarded GF $G_{0}(A|B)_{\omega+i\eta}$. However, $G_{1}(A|B)_{\omega}$ has real second-order poles and thus violates the causality. Moreover, some terms in $G_{1}(A|B)_{\omega}$, including $Z_1 = - \beta \sum_{m} E_{m}^{(1)}e^{-\beta E_{m}^{(0)}}$, have a factor $\beta$ which comes from the Taylor expansion of $E_m$ in the Boltzmann factor. Thus $G_{1}(A|B)_{\omega}$ will diverge in the zero temperature limit unless $\beta$ factors in different terms cancel. We call this problem the zero-temperature divergence problem, which occurs when the ground state of $H$ is unanalytical at $\lambda=0$. In general, the $k$-th order term $G_{k}(A|B)_{\omega}$ has $(k+1)$-th order poles and contains terms with factor $\beta^{k}$. As a result, a truncated series of GF almost always violates the analytical structure and diverges at zero temperature. In order to avoid these problems, it is necessary to sum part of the expansion contributions to infinite order.

In the following, we discuss possible resummation methods to recover the analytical structure of GF. The zero-temperature divergence problem will be considered in the next subsection, invoking the self-consistent EOM expansion. For interacting electron systems, a conventional practice is to directly expand the self-energy (SE) up to some finite order and then insert it into the Dyson equation to produce GF. Suppose we have obtained the series expansion of the single-particle GF $G(c_{k\sigma}|c_{k\sigma}^{\dagger})_{\omega}$ for a lattice Hamitonian up to order $\lambda^n$
\begin{eqnarray}   \label{eq:18}
 &&   G(c_{k\sigma}|c_{k\sigma}^{\dagger})_{\omega} \nonumber \\
 &\approx &  G_{0}(c_{k\sigma}|c_{k \sigma}^{\dagger})_{\omega} +  \lambda G_{1}(c_{k\sigma}|c_{k \sigma}^{\dagger})_{\omega} + ... +  \lambda^{n} G_{n}(c_{k \sigma}|c_{k \sigma}^{\dagger})_{\omega}. \nonumber \\
&& 
\end{eqnarray}
Here $c_{k \sigma}$ and $c_{k \sigma}^{\dagger}$ are electron creation and annihilation operators, respectively. The GF obtained from the SE resummation is denoted as  $\overline{G}_{SE}(c_{k \sigma}|c_{k \sigma}^{\dagger})_{\omega}$,
\begin{equation}   \label{eq:19}
   \overline{G}_{SE}(c_{k \sigma}|c_{k \sigma}^{\dagger})_{\omega} = \frac{1}{\mathcal{G}_{0}^{-1}(k, \omega)  - \Sigma(c_{k \sigma}|c_{k \sigma}^{\dagger})_{\omega} }.
\end{equation}
Here, $\mathcal{G}_{0}^{-1}(k, \omega) = \omega + \mu - \epsilon_{k}$ is the exact GF of the non-interacting Hamiltonian $H(V=0)$. Here $V$ represents the interaction strength in $H$. In Eq.(\ref{eq:19}), we substitute the SE with its formal series truncated at $n$-th order $\Sigma(c_{k \sigma}|c_{k \sigma}^{\dagger})_{\omega} \approx \Sigma_0(c_{k \sigma}|c_{k \sigma}^{\dagger})_{\omega} + \lambda \Sigma_1(c_{k \sigma}|c_{k \sigma}^{\dagger})_{\omega} + ... + \lambda^n \Sigma_{n}(c_{k \sigma}|c_{k \sigma}^{\dagger})_{\omega}$. 
$\mathcal{G}_{0}^{-1}(k, \omega)$ should also be expanded if $H(V=0)$ depends on $\lambda$. Expanding $\overline{G}_{SE}(c_{k \sigma}|c_{k \sigma}^{\dagger})_{\omega}$ into power series of $\lambda$ and comparing it with the series of $G(c_{k\sigma}|c_{k\sigma}^{\dagger})_{\omega}$ to order $\lambda^n$, we can fix the expansion terms of SE. $\overline{G}_{SE}(c_{k \sigma}|c_{k \sigma}^{\dagger})_{\omega}$ obtained in this way is exact up to order $\lambda^n$ but contains approximate terms from $\lambda^{n+1}$ to $\lambda^{\infty}$. Note that other equations derived from the diagramatic resummation, such as the Larkin equation,\cite{Sherman1} could also be used to do resummation in a similar way.

In the strong-coupling expansion of the Anderson impurity model, the SE resummation method has been used to calculate the impurity GF.\cite{Dai1} The resulting GF violates the causality and does not obey the sum rule (see below). Similar problems such as negative spectral function also appear in the weak-coupling expansions.\cite{Stefanucci1} Inspecting the analytical structure of SE $\Sigma(c_{k \sigma}|c_{k \sigma}^{\dagger})_{\omega} = c_{0k} + \sum_{m} c_{mk}/(\omega - \omega_{mk})$,\cite{Luttinger1} we find that a truncated series of SE violates the correct analytical structure and has zero-temperature divergence problem, if only the poles $\{ \omega_{mk} \}$ and weights $\{ c_{mk}\}$ contain $\lambda$ and $\beta$. In special cases where the poles have $\lambda^{i \geq n}$ contributions only, the SE expansion to order $\lambda^{n-1}$ does not produce non-simple poles and the causality will be fulfilled. As will be shown below, this is the case of the weak-coupling expansion to order $U^2$. We conclude that in general, the resummation method from the truncated bare expansion of SE can guarantee neither the correct analytical structure nor the correct zero temperature limit.

To overcome the problem with analytical structure, Pairault {\it et al.} suggested a resummation method based on the continued fraction (CF).\cite{Pairault1,Pairault2} For the single particle GF $G(c_{k \sigma}|c_{k \sigma}^{\dagger})_{\omega}$, one can construct a CF of the form
\begin{eqnarray}   \label{eq:20}
&& \overline{G}_{CF}(d_{\sigma}|d_{\sigma}^{\dagger})_{\omega}  \nonumber \\
&& = \cfrac{a_0(\lambda)}{\omega + b_1(\lambda) 
          - \cfrac{a_1(\lambda)}{\omega + b_2(\lambda)
          - \cfrac{a_2(\lambda)}{\omega + b_3(\lambda) -... } } }. \nonumber \\
  &&
\end{eqnarray}
It was proven that for real parameters $a_l(\lambda) \geq 0$ and real $b_l(\lambda)$ ($l=1,2,...$), the above expression always gives the correct analytical structure of GF, i.e., it consists of real simple poles. This can be understood by regarding Eq.(\ref{eq:20}) as the local GF of a free semi-infinite chain Hamiltonian. To carry out the CF resummation, we formally expand the coefficients $a_l(\lambda)$ and $b_l(\lambda)$ ($l=1,2,...$) into Taylor series of $\lambda$ up to $\lambda^{n}$ order, and then compare the obtained $G_{CF}(d_{\sigma}|d_{\sigma}^{\dagger})_{\omega}$ with $G(d_{\sigma}|d_{\sigma}^{\dagger})_{\omega}$ to order $\lambda^n$ to fix the expansion coefficients of $a_l(\lambda)$ and $b_l(\lambda)$. Usually, due to the finite number of poles produced by the GF expansion, a  CF with finite levels is sufficient for this task. Besides being exact up to order $\lambda^n$, the obtained GF $\overline{G}_{CF}(d_{\sigma}|d_{\sigma}^{\dagger})_{\omega}$ is causal. This resummation method was used in the strong-coupling expansion study of two dimensional Hubbard model.\cite{Pairault1,Pairault2,Brune1}

\subsection{Self-Consistent EOM Series Expansion}
The CF resummation method can overcome the causality problem in the truncated series of GF and recover the correct analytical structure. However, it still has the problem of zero-temperature divergence. Using Lehmannn representation, we have shown that in general, $\beta$ factors will appear in $G_{i}(A|B)_{\omega}$ ($i \geq 1$). The CF resummation procedure transmits these $\beta$ factors into the parameters $a_l(\lambda)$ and $b_l(\lambda)$, leading to an unphysical shifting of certain poles to infinity as $T$ approaches zero. Indeed, in the strong-coupling expansion study of a two dimensional Hubbard model,\cite{Pairault1,Pairault2,Brune1,Dupuis1} $\beta$ factors explicitly appear both in the bare expansion terms of GF and in the parameters of CF. The validity range of the CF-resummed GF is thus limited to temperatures higher than some energy scale. For the Hubbard model, this effect was attributed to the fact that the intra-Hubbard-band hopping process of electrons involves the energy scale $t/T$ instead of $t/U$\cite{Dupuis1} and cannot be described accurately by $t/U$ expansion. Generally speaking, this problem reflects that the ground state energy or the density matrix is not expanded on the same footing as the excitation energies. Inspecting the Lehmann representation of SE shows that this problem also exists in the SE resummation. 

In the bare expansion formalism Eq.(\ref{eq:9}), the only place where $\beta^i$ emerges is the component $\langle [A, B]_{\pm}\rangle_i$ ($i \ge 1$). $G_{i}(A|B)_{\omega}$ generally has ($i+1$)-th order poles. When $\langle [A, B]_{\pm}\rangle_i$ ($i \ge 1$) is calculated from the same order GF component using the fluctuation-dissipation theorem, $\beta^{i}$ will be produced through the frequency integration. Therefore, this problem could be solved if the averages are not calculated order by order from each $G_{i}(A|B)_{\omega}$, but from the full $G(A|B)_{\omega}$ after its correct analytical structure has been restored. Below, we propose the self-consistent GF EOM expansion scheme in this spirit.

We start from the EOM for the full GF Eq.(\ref{eq:3}). For simplicity, here we consider the left-side EOM only. The corresponding formula for the right-side EOM can be derived similarly. We define the renormalized zeroth-order GF $G_{0}(A|B)_{\omega}$ by the EOM
\begin{equation}   \label{eq:21}
 \omega G_0(A|B)_{\omega} = \langle \left[A, B \right]_{\pm} \rangle + G_0([A, H_0]|B)_{\omega} .
\end{equation}
Unlike in the bare expansion Eq.(\ref{eq:10}), $\langle \left[A, B \right]_{\pm} \rangle$ here is the thermodynamical average with respect to full Hamiltonian $H$. It will be calculated self-consistently from the CF-resummed GF which has correct analytical structure. Subtracting Eq.(\ref{eq:21}) from the the left-side EOM Eq.(\ref{eq:3}) and defining the renormalized $n$-th order residue as $\Gamma_{n}(A|B)_{\omega} \equiv G(A|B)_{\omega}-G_0(A|B)_{\omega} - G_1(A|B)_{\omega} -... - G_n(A|B)_{\omega} $, we get the EOM for the zeroth-order residue $\Gamma_0(A|B)_{\omega}$ as
\begin{equation}   \label{eq:22}
 \omega \Gamma_0(A|B)_{\omega} = G_0([A, H_1]|B)_{\omega} + \Gamma_0([A,H]|B)_{\omega}.
\end{equation}
For the next order renormalized GF $G_{1}(A|B)_{\omega}$, we require that it satisfy the EOM of $\Gamma_0(A|B)_{\omega}$ at the leading order of $\lambda$, i.e., Eq.(\ref{eq:22}) with $H$ replaced by $H_0$. We have 
\begin{equation}   \label{eq:23}
 \omega G_1(A|B)_{\omega} = G_0([A, H_1]|B)_{\omega} + G_1([A,H_0]|B)_{\omega}.
\end{equation}
Similarly, EOM of the residue $\Gamma_1(A|B)_{\omega} = \Gamma_0(A|B)_{\omega}-G_1(A|B)_{\omega} $ can be obtained by subtracting Eq.(\ref{eq:23}) from Eq.(\ref{eq:22}). This procedure is carried out repeatedly to produce EOM of the renormalized GF component for $i \geq 1$,
\begin{equation}   \label{eq:24}
 \omega G_i(A|B)_{\omega} = G_{i-1}([A, H_1]|B)_{\omega} + G_i([A,H_0]|B)_{\omega}. 
\end{equation}
For $i=0$, Eq.(\ref{eq:21}) applies.
The $n$-th order residue $\Gamma_n(A|B)_{\omega}$ satisfies the EOM
\begin{equation}   \label{eq:25}
 \omega \Gamma_n(A|B)_{\omega} = G_{n}([A, H_1]|B)_{\omega} + \Gamma_n([A,H]|B)_{\omega}. 
\end{equation}
Series expansion using the right-hand side EOM can be derived similarly.
Note that in this self-consistent series expansion, one can use either the left-hand side or the right-hand side EOM formula throughout the derivation,  but cannot mix them in different orders. This differs from the bare EOM expansion.

This EOM expansion scheme only involves full averages $\langle \left[A, B \right]_{\pm} \rangle$ which need to be calculated self-consistently from the corresponding full GFs.
To be free from the zero-temperature divergence problem, these GF must have real simple poles and therefore CF-resummed GF should be used. 

A formal solution of the above self-consistent EOM expansion can be obtained in terms of the Liouville operator $\mathcal{L}$ as,
\begin{eqnarray}   \label{eq:26}
&&   G_{0}(A|B)_{\omega} = \Big\langle \left\{\frac{1}{\omega - \mathcal{L}_{0}}A, B \right\} \Big\rangle ; \nonumber \\
&&   G_{i}(A|B)_{\omega} = G_{i-1}\left(\mathcal{L}_{1} \frac{1}{\omega - \mathcal{L}_{0}}A \Big|B \right)_{\omega} \,\,\,(i \geq 1);   \nonumber \\
&&   \Gamma_{n}(A|B)_{\omega} = G_{n}\left( \mathcal{L}_{1} \frac{1}{\omega - \mathcal{L}}A  \Big| B \right)_{\omega} .
\end{eqnarray}
Here, $\mathcal{L}_{0}$ and $\mathcal{L}_{1}$ are the Liouville operators of $H_0$ and $H_1$, respectively. They act on any operator $\hat{O}$ as $\mathcal{L}_{0} \hat{O} = [ \hat{O}, H_0 ]$ and $\mathcal{L}_{1} \hat{O} =[ \hat{O}, H_1] $.

\section{Weak-Coupling Expansion for Anderson Impurity Model}

In this section, we apply the formalism developed above to the single impurity Anderson model. This model is one of the best studied models for correlated electron systems, due to its importance in the dilute magnetic impurity problem, in the quantum dot physics, as well as in the application of dynamical mean-field theory for Hubbard model. The Hamiltonian of the single impurity Anderson model reads
\begin{eqnarray}    \label{eq:27}
   H_{Aim} &=& \sum_{k \sigma} (\epsilon_{k\sigma}-\mu) c_{k \sigma}^{\dagger} c_{k\sigma} + \sum_{k \sigma } V_{k\sigma} \left(c_{k \sigma}^{\dagger} d_{\sigma} + d_{\sigma}^{\dagger} c_{k \sigma} \right)  \nonumber \\
   &+& (\epsilon_d - \mu) \sum_{\sigma} n_{\sigma} + U n_{\uparrow} n_{\downarrow}.
\end{eqnarray}
Here we consider spin-dependent energies and hybridizations of bath electrons. $n_\sigma = d_{\sigma}^{\dagger}d_{\sigma}$ is the electron number operator of impurity. The influence of bath to impurity is described by the hybridization function $\Delta_{\sigma}(\omega) = \sum_k V^2_{k \sigma} \delta(\epsilon - \epsilon_{k \sigma})$. Throughout this work, we set the chemical potential $\mu=0$ as the zero point of frequency.

Both the weak- and the strong-coupling expansion for this model have been obtained before. Here, for demonstration purpose only, we apply the EOM series expansion method to derive the weak-coupling expansion for the local GF to $U^2$ order, to recover the well-known results of Yamada\cite{Yamada1} at the same level.

In the weak-coupling expansion, we decompose $H_{Aim} = H_0 + H_1$ as
\begin{equation}    \label{eq:28}
   H_0 = \sum_{k \sigma} \epsilon_{k\sigma} c_{k \sigma}^{\dagger} c_{k\sigma} + \sum_{k \sigma } V_{k\sigma} \left(c_{k \sigma}^{\dagger} d_{\sigma} + d_{\sigma}^{\dagger} c_{k \sigma} \right) +  \sum_{\sigma} \tilde{\epsilon}_{d \sigma} n_{\sigma},
\end{equation}
and 
\begin{equation}    \label{eq:29}
   H_{1} =  U n_{\uparrow} n_{\downarrow} - \sum_{\sigma} \alpha_{\sigma} n_{\sigma}.
\end{equation}
Here, $\tilde{\epsilon}_{d \sigma} = \epsilon_d + \alpha_{\sigma}$. $\alpha_{\sigma}$ is a parameter to be determined by the principle of convenience. For an example, its value could be fixed by requiring that the first-order contribution $G_{1}(d_{\sigma}|d_{\sigma}^{\dagger})_{\omega}$ is zero. In that case, expanding GF into Taylors series of $H_1$ amounts to perturbation around the Hartree-Fock Hamiltonian.

To facilitate comparison, we will apply the bare EOM series expansion and use the SE resummation. Here the GF is of fermion type and the anti-commutator of operators $A$ and $B$ is denoted as $\{A, B \}$. The zeroth-order GF $G_{0}(d_{\sigma}|d_{\sigma}^{\dagger})_{\omega}$ is easily solved from its EOM,
\begin{equation}    \label{eq:30}
   G_{0}(d_{\sigma}|d_{\sigma}^{\dagger})_{\omega} = \frac{1}{\omega - \tilde{\epsilon}_{d \sigma} - \Gamma_{\sigma}(\omega)}.
\end{equation}
Here $\Gamma_{\sigma}(\omega) = \sum_{k} V^{2}_{k \sigma}/(\omega -\epsilon_{k\sigma})$.

For $i \geq 1$, the EOM for $G_{i}(d_{\sigma}|d_{\sigma}^{\dagger})_{\omega}$ reads
\begin{eqnarray}    \label{eq:31}
  && \omega G_{i}(d_{\sigma}|d_{\sigma}^{\dagger})_{\omega}  \nonumber \\
  &&= \langle \{d_{\sigma}, d_{\sigma}^{\dagger} \} \rangle_{i} + G_{i-1}( [d_{\sigma}, H_1]|d_{\sigma}^{\dagger})_{\omega} + G_{i}( [d_{\sigma}, H_0]|d_{\sigma}^{\dagger})_{\omega}. \nonumber \\
&&  
\end{eqnarray}
Using $\langle 1 \rangle_{i \geq 1} = 0$, $[d_{\sigma}, H_0] = \sum_k V_k c_{k \sigma} + \tilde{\epsilon}_{d\sigma} d_{\sigma} $ and $[d_{\sigma}, H_1] = U n_{\bar{\sigma}}d_{\sigma} - \alpha_{\sigma} d_{\sigma}$, we get  
\begin{eqnarray}    \label{eq:32}
&&   (\omega - \tilde{\epsilon}_{d\sigma}) G_{i}(d_{\sigma}|d_{\sigma}^{\dagger})_{\omega} \nonumber \\
&=&  \sum_{k} V_{k} G_{i}(c_{k\sigma}|d_{\sigma}^{\dagger})_{\omega} - \alpha_{\sigma} G_{i-1}(d_{\sigma}|d_{\sigma}^{\dagger})_{\omega} 
\nonumber \\
&& + U G_{i-1}(n_{\bar{\sigma}}d_{\sigma}|d_{\sigma}^{\dagger})_{\omega}.
\end{eqnarray}
We use the left-side EOM for the $i$-th order new GF $G_{i}(c_{k\sigma}|d_{\sigma}^{\dagger})_{\omega}$ to obtain
\begin{equation}    \label{eq:33}
  G_{i}(c_{k\sigma}|d_{\sigma}^{\dagger})_{\omega} = \frac{V_{k\sigma}}{\omega - \epsilon_{k\sigma}} G_{i}(d_{\sigma}|d_{\sigma}^{\dagger})_{\omega}.
\end{equation}
Putting Eq.(\ref{eq:33}) into Eq.(\ref{eq:32}), we get
\begin{eqnarray}    \label{eq:34}
&&  G_{i}(d_{\sigma}|d_{\sigma}^{\dagger})_{\omega}  \nonumber \\
&=& G_{0}(d_{\sigma}| d_{\sigma}^{\dagger} )_{\omega}  \left[ U G_{i-1}(  n_{\bar{\sigma}} d_{\sigma}| d_{\sigma}^{\dagger} )_{\omega} - \alpha_{\sigma}G_{i-1}( d_{\sigma}| d_{\sigma}^{\dagger} )_{\omega} \right]. \nonumber \\
&&
\end{eqnarray}
For $i=1$, this equation involves $G_{0}(n_{\bar{\sigma}}d_{\sigma}|d_{\sigma}^{\dagger})_{\omega}$, which is easily solved from its right-side EOM as $G_{0}(n_{\bar{\sigma}}d_{\sigma}|d_{\sigma}^{\dagger})_{\omega} = \langle n_{\bar{\sigma}} \rangle_{0} G_{0}(d_{\sigma}|d_{\sigma}^{\dagger})_{\omega}$. Eq.(\ref{eq:34}) then produces
\begin{equation}     \label{eq:35}
   G_{1}(d_{\sigma}|d_{\sigma}^{\dagger})_{\omega} =  \frac{U \langle  n_{\bar{\sigma}} \rangle_{0} - \alpha_{\sigma} }{ \left[ \omega - \tilde{\epsilon}_{d \sigma}- \Gamma_{\sigma}(\omega)\right]^2}.
\end{equation}
For $i=2$, Eq.(\ref{eq:34}) involves a new GF $G_{1} (n_{\bar{\sigma}}d_{\sigma}|d_{\sigma}^{\dagger})_{\omega}$. After some calculation, its EOM is solved to produce 
\begin{eqnarray}    \label{eq:36}
&&  G_{1}(n_{\bar{\sigma}}d_{\sigma}|d_{\sigma}^{\dagger})_{\omega} \nonumber \\
&&= \left[ \langle n_{\bar{\sigma}} \rangle_{1} +  U K_{\sigma}(\omega) \right] G_{0}(d_{\sigma}| d_{\sigma}^{\dagger} )_{\omega}  \nonumber \\
&&+ \langle n_{\bar{\sigma}} \rangle_{0} \left[ U \langle n_{\bar{\sigma}} \rangle_{0} - \alpha_{\sigma}\right] G_{0}^{2}(d_{\sigma}| d_{\sigma}^{\dagger} )_{\omega}. \nonumber \\
&&
\end{eqnarray}
Details of this calculation is presented in Appendix A. Putting Eq.(\ref{eq:35}) and Eq.(\ref{eq:36}) into Eq.(\ref{eq:34}) we obtain
\begin{eqnarray}    \label{eq:37}
 G_{2}(d_{\sigma}|d_{\sigma}^{\dagger})_{\omega} &=&
U \left[\langle n_{\bar{\sigma}} \rangle_{1} + U K_{\sigma}(\omega) \right] G_{0}^{2}(d_{\sigma}| d_{\sigma}^{\dagger} )_{\omega}   \nonumber \\
&& + \left[U \langle n_{\bar{\sigma}} \rangle_{0} - \alpha_{\sigma} \right]^2 
G_{0}^{3}(d_{\sigma}| d_{\sigma}^{\dagger} )_{\omega}. \nonumber \\
&&
\end{eqnarray}
The function $K_{\sigma}(\omega)$ appearing here reads
\begin{eqnarray}    \label{eq:38}
&& K_{\sigma}(\omega)   \nonumber \\
&=& \displaystyle\iiint_{-\infty}^{\infty}  \frac{ \rho_{0 \bar{\sigma}}(\epsilon_1) \rho_{0 \bar{\sigma}}(\epsilon_2) \rho_{0 \sigma}(\epsilon_3) }{\omega + \epsilon_1 - \epsilon_2 -\epsilon_3} F(\epsilon_1, \epsilon_2, \epsilon_3) d\epsilon_1 d\epsilon_2 d\epsilon_3,  \nonumber \\
&& 
\end{eqnarray}
where $\rho_{0 \sigma}(\epsilon) = -1/\pi {\text Im} G_{0}(d_{\sigma}|d_{\sigma}^{\dagger})_{\epsilon + i\eta}$ is the free local density of states  of spin $\sigma$ and 
\begin{equation}    \label{eq:39}
  F(\epsilon_1, \epsilon_2, \epsilon_3) = n(\epsilon_3) \left[n(\epsilon_2)-n(\epsilon_1) \right] + n(\epsilon_1)\left[ 1-n(\epsilon_2) \right],
\end{equation}
with $n(\epsilon) = 1/(e^{\beta \epsilon} + 1)$ being the Fermi-Dirac distribution function.
The unknown averages involved in the above expressions are $\langle n_{\bar{\sigma}} \rangle_{0}$ and $\langle n_{\bar{\sigma}} \rangle_{1}$. From 
\begin{equation}    \label{eq:40}
 \langle n_{\bar{\sigma}} \rangle_{i} = -\frac{1}{\pi} \int_{-\infty}^{\infty} 
 {\text Im}  G_{i}(d_{\bar{\sigma}}|d_{\bar{\sigma}}^{\dagger})_{\omega+ i\eta} \frac{1}{e^{\beta \omega} + 1} d\omega.
\end{equation}
they are calculated as
\begin{eqnarray}    \label{eq:41}
  \langle n_{\bar{\sigma}} \rangle_{0} &=& \displaystyle\int_{-\infty}^{\infty} \rho_{0 \bar{\sigma}} (\epsilon) n(\epsilon) d\epsilon ;    \nonumber  \\
  \langle n_{\bar{\sigma}} \rangle_{1} &=& \left( U \langle n_{\sigma} \rangle_{0} - \alpha_{\bar{\sigma}} \right)  \times  \nonumber \\
&& \displaystyle\int_{-\infty}^{\infty} \frac{\rho_{0 \bar{\sigma}}(\epsilon_1)  \rho_{0 \bar{\sigma}}(\epsilon_2) }{ \epsilon_{1} - \epsilon_{2}} \left[ n(\epsilon_1) - n(\epsilon_2) \right] d\epsilon_1 d \epsilon_2.  \nonumber \\
&&
\end{eqnarray}

Until now, we have obtained the lowest three orders of GF, $G_{i}(d_{\sigma}|d_{\sigma}^{\dagger})_{\omega}$ ($i=0,1,2$). To carry out the SE resummation, we insert the truncated expansion of SE $\Sigma_{\sigma}(\omega) \approx \Sigma_{0 \sigma}(\omega) + \lambda \Sigma_{1 \sigma}(\omega) + \lambda^2 \Sigma_{2 \sigma}(\omega)$ into the Dyson equation $G_{SE}(d_{\sigma}|d_{\sigma}^{\dagger})_{\omega} = \left[ \mathcal{G}_{0 \sigma}^{-1}(\omega)  - \Sigma_{\sigma}(\omega) \right]^{-1}$, expand it into a Taylor series of $\lambda$, and compare the $\lambda^i$ term with $G_{i}(d_{\bar{\sigma}}|d_{\bar{\sigma}}^{\dagger})_{\omega}$ ($i=0,1,2$). $\mathcal{G}_{0 \sigma}(\omega) = 1/\left[\omega - \epsilon_d - \Gamma_{\sigma}(\omega) \right]$ is the non-interacting local GF. The SE is given as 
\begin{eqnarray}    \label{eq:42}
 \Sigma_{0 \sigma}(\omega) &=& \alpha_{\sigma};  \nonumber \\
 \Sigma_{1 \sigma}(\omega) &=& G_{0}^{-2}(d_{\sigma}|d_{\sigma}^{\dagger})_{\omega} G_{1}(d_{\sigma}|d_{\sigma}^{\dagger})_{\omega} ;  \nonumber \\
 \Sigma_{2 \sigma}(\omega) &=&   G_{0}^{-2}(d_{\sigma}|d_{\sigma}^{\dagger})_{\omega} G_{2}(d_{\sigma}|d_{\sigma}^{\dagger})_{\omega}  \nonumber \\
  && - G_{0}^{-3}(d_{\sigma}|d_{\sigma}^{\dagger})_{\omega} G_{1}^{2}(d_{\sigma}|d_{\sigma}^{\dagger})_{\omega}. 
\end{eqnarray}

Now we can choose $\alpha_{\sigma}$ to simplify the expression. By assigning $\alpha_{\sigma} = U\langle n_{\bar{\sigma}}\rangle_{0}$, the expansion gets simplified as
\begin{eqnarray}    \label{eq:43}
&& G_{0}(d_{\sigma}|d_{\sigma}^{\dagger})_{\omega} = \left[\omega - \tilde{\epsilon}_{d \sigma} - \Gamma_{\sigma}(\omega) \right]^{-1} ; \nonumber \\
&& G_{1}(d_{\sigma}|d_{\sigma}^{\dagger})_{\omega} =0;  \nonumber \\
&&  G_{2}(d_{\sigma}|d_{\sigma}^{\dagger})_{\omega}= G_{0}^{2}(d_{\sigma}| d_{\sigma}^{\dagger} )_{\omega}\left[ U \langle n_{\bar{\sigma}} \rangle_{1}   +  U^2 K_{\sigma}(\omega) \right].  \nonumber \\
&&
\end{eqnarray}
The corresponding SE reads
\begin{equation}    \label{eq:44}
   \Sigma_{\sigma}(\omega) = U \left[ \langle n_{\bar{\sigma}} \rangle_{0} + \langle n_{\bar{\sigma}} \rangle_{1} \right] +  U^2 K_{\sigma}(\omega) + O(U^3).
\end{equation}
The SE-resummed GF is finally obtained as $G_{SE}(d_{\sigma}|d_{\sigma}^{\dagger})_{\omega} = \left[ \mathcal{G}_{0 \sigma}^{-1}(\omega)  - \Sigma_{\sigma}(\omega) \right]^{-1}$. 
Note that $\langle n_{\bar{\sigma}} \rangle_{1}$ has a factor $\beta$ and may lead to zero-temperature divergence in SE. Here, the causality problem does not appear since the obtained SE has real simple poles only. The reason, as analysed in general in Sec.IIB, is that the poles of SE for the Anderson impurity model contain terms of the order $U^{i \geq 3}$ and expansion to order $U^2$ does not produce non-simple poles. In the paramagnetic and particle-hole symmetric case, $\langle n_{\bar{\sigma}} \rangle_{0} =1/2$ and $\langle n_{\bar{\sigma}} \rangle_{1} =0$, the zero-temperature divergence problem does not appear. The above expression is equivalent to the Matsubara SE obtained by Yamada,\cite{Yamada1}
\begin{eqnarray}    \label{eq:45}
&&   \Sigma_{\sigma}(i \omega_n) = \frac{U}{2} +  U^2 \displaystyle\int_{0}^{\beta} G_{0}^{3}(\tau) e^{i \omega_n \tau} d\tau;   \nonumber \\
&&  G_{0}(i\omega_n) = \left[i \omega_n - \Gamma_{\sigma}(i\omega_n) \right]^{-1}. 
\end{eqnarray}

When combined with the dynamical mean-field theory, the above SE produces the iterative perturbation theory\cite{Yamada1} (IPT) which describes the Mott metal-insulator transition very well. For the particle-hole symmetric case and paramagnetic bath, the above bare expansion with SE resummation has neither causality problem nor zero-temperature divergence problem due to $\langle n_{\overline{\sigma}} \rangle= 1/2$. This simplest form of IPT fails, however, away from particle-hole symmetry or in the magnetic bath. Extension of the original IPT to such situations received some research effort \cite{Kajueter1} in the spirit of interpolation between various exact limits. Eq.(\ref{eq:44}) reminds us that the zero-temperature divergence problem may occur in the bare expansion. The recovery of atomic limit is also difficult by SE-resummation away from particle-hole symmetry. These problems could be remedied by the self-consistent EOM expansion supplemented with CF resummation. For the moment, we leave in depth discussions of this issue to the future and focus on the strong-coupling expansion of the GF.

\section{Strong-Coupling Expansion for Anderson Impurity Model}

In this section, we carry out the strong-coupling expansion for $G(d_{\sigma}|d_{\sigma}^{\dagger})_{\omega}$ to the order $V_{k}^2$. 
Besides testing the EOM expansion methods, the obtained formula may also serve as a useful strong-coupling impurity solver for the dynamical mean-field theory to describe the anti-ferromagnetic insulating phase, for which existing weak-coupling-based theories such as IPT and the functional renormalization group method\cite{Karrasch1} are faced with difficulties.

We investigate the impurity spectral function, electron occupation, and the double occupancy at the particle-hole symmetric point both for the paramagnetic and the magnetic cases. In the paramagnetic case, our results agree with the direct expansion results.\cite{Dai1} We compare the local density of states from three different calculation schemes: bare EOM expansion supplemented with SE resummation, bare EOM expansion with CF resummation, and the self-consistent expansion with CF resummation.

To do the strong-coupling expansion, we decompose $H_{Aim}= H_0 + H_1$ into
\begin{equation}    \label{eq:46}
   H_0 = \sum_{k \sigma} \epsilon_{k\sigma} c_{k \sigma}^{\dagger} c_{k\sigma} +  U n_{\uparrow} n_{\downarrow} + \epsilon_d \sum_{\sigma} n_{\sigma},
\end{equation}
and 
\begin{equation}    \label{eq:47}
   H_{1} =\sum_{k \sigma } V_{k\sigma} \left(c_{k \sigma}^{\dagger} d_{\sigma} + d_{\sigma}^{\dagger} c_{k \sigma} \right).
\end{equation}
Obviously, $H_0$ is exactly solvable. Due to the existence of $Un_{\uparrow}n_{\downarrow}$ in $H_0$, the hierarchy EOM for the local GF closes at the second level and it is more convenient to work with SBO formalism. We denote the eigen-state and eigen-energy of $\hat{h}_0 \equiv U n_{\uparrow} n_{\downarrow} + \epsilon_d \sum_{\sigma} n_{\sigma}$ as $|\alpha \rangle$ and $E_{\alpha}$, respectively. That is, $\hat{h}_0 |\alpha \rangle = E_{\alpha} |\alpha \rangle$. The SBOs $\{ A_{\alpha \beta} \}$ are defined as the projector operators $A_{\alpha \beta} \equiv |\alpha \rangle \langle \beta|$. They satisfy the algebraic relations $A_{\alpha \beta} A_{\mu \nu} = \delta_{\beta \mu} A_{\alpha \nu}$ and $\sum_{\alpha} A_{\alpha \alpha}=1$. The latter plays the role of the kinematic sum rule\cite{Callen1} and is important for the self-consistent solution for the averages. $H_0$ is now written as
\begin{equation}    \label{eq:48}
   H_0 = \sum_{k \sigma} \epsilon_{k\sigma} c_{k \sigma}^{\dagger} c_{k\sigma} +  \sum_{\mu} E_{\mu} |\mu \rangle \langle \mu |.
\end{equation}
We express the operator $d_{\sigma}$ in $H_{1}$ as $d_{\sigma} = \sum_{\mu\nu} f^{\sigma}_{\mu \nu} A^{\sigma}_{\mu\nu}$ and $f^{\sigma}_{\mu \nu} = \langle \mu |d_{\sigma} | \nu \rangle$. Here the superscript $\sigma$ in $A^{\sigma}_{\mu\nu}$ denotes that this SBO, when acting on a state, decreases the number of spin-$\sigma$ electrons by $1$. It is a Grassmann odd operator. We use $A_{\mu\nu}$ without the superscript for general SBOs with unspecified quantum numbers. For Grassmann even (odd) $A_{\alpha \beta}$, its commutator (anticommutator) with $c_{k\sigma}$ is zero. 
 The SBO formalism has been used in the study of the Heisenberg model with large spin.\cite{Haley2} Recently this formalism is employed to combine the GF EOM truncation approximation with the exact diagonalization method to develop a new impurity solver for the dynamical mean-field theory.\cite{Li1}

\subsection{Bare Strong-Coupling Expansion to $V^2_k$ Order}

Below, we first carry out the bare EOM series expansion for $G(A^{\sigma }_{\alpha \beta}|A_{\gamma \delta}^{\sigma \dagger})_{\omega}$. The local single particle GF is given by $G(d_{\sigma}|d_{\sigma}^{\dagger})_{\omega} = \sum_{\alpha\beta}\sum_{\gamma \delta} f^{\sigma}_{\alpha\beta} f^{\sigma *}_{\gamma\delta} G(A^{\sigma }_{\alpha \beta}|A_{\gamma \delta}^{\sigma  \dagger})_{\omega}$. The EOM of the zeroth-order GF reads
\begin{equation}    \label{eq:49}
   \omega G_{0}(A^{\sigma}_{\alpha \beta}| A_{\gamma \delta} ^{\sigma \dagger} )_{\omega} = 
  \langle \{A^{\sigma}_{\alpha \beta}, A_{\gamma \delta}^{\sigma \dagger} \} \rangle_0 + G_{0}(\left[ A^{\sigma}_{\alpha \beta}, H_0 \right]|A_{\gamma \delta}^{\sigma \dagger})_{\omega}.
\end{equation}
Using the relations $\{A^{\sigma}_{\alpha \beta}, A_{\gamma \delta}^{\sigma \dagger} \} = \delta_{\beta \delta} A_{\alpha \gamma} + \delta_{\gamma \alpha} A_{\delta\beta}$, $\langle A_{\alpha\beta}\rangle_{0} = \delta_{\alpha\beta} e^{-\beta E_{\alpha}}/Z_0$ and $\left[ A^{\sigma }_{\alpha \beta}, H_0 \right] = \left(E_{\beta} - E_{\alpha} \right) A^{\sigma }_{\alpha \beta}$, we obtain
\begin{equation}    \label{eq:50}
   G_{0}(A^{\sigma}_{\alpha \beta}|A_{\gamma \delta}^{\sigma \dagger})_{\omega} = 
\delta_{\alpha\gamma} \delta_{\beta \delta} \frac{a_{\alpha} + a_{\beta}}{\omega + E_{\alpha} -E_{\beta}}.
\end{equation}
Here $a_{\alpha} = e^{-\beta E_{\alpha}}/Z_0$, and $Z_0 = \sum_{\mu} e^{-\beta E_{\mu}}$ is the partition function of $H_0$.

For $i \geq 1$, the EOM reads
\begin{eqnarray}    \label{eq:51}
 \omega G_{i}(A^{\sigma}_{\alpha \beta}|A_{\gamma \delta} ^{\sigma \dagger})_{\omega}  &=&  \langle \{A^{\sigma}_{\alpha \beta}, A_{\gamma \delta}^{\sigma \dagger} \} \rangle_{i} + G_{i-1}(\left[ A^{\sigma}_{\alpha \beta}, H_1 \right]|A_{\gamma \delta}^{\sigma \dagger} )_{\omega}   \nonumber \\
 && + G_{i}(\left[ A^{\sigma}_{\alpha \beta}, H_0 \right]|A_{\gamma \delta}^{\sigma \dagger} )_{\omega}.
\end{eqnarray}
To simplify the notation, we expand the anti-commutators involved in $ \left[ A^{\sigma}_{\alpha \beta}, H_1 \right]$ as
\begin{eqnarray}    \label{eq:52}
&&  \{ A^{\sigma}_{\alpha \beta}, d_{\sigma^{\prime}}^{\dagger} \} = \sum_{\mu\nu} M^{\sigma\sigma'}_{\alpha\beta, \mu\nu} A_{\mu\nu}; \nonumber \\
&&  \{ A^{\sigma}_{\alpha \beta}, d_{\sigma^{\prime}} \} = \sum_{\mu\nu} N^{\sigma\sigma'}_{\alpha\beta, \mu\nu} A_{\mu\nu}.
\end{eqnarray}
The coefficients $M^{\sigma\sigma'}_{\alpha\beta, \mu\nu}$ and $N^{\sigma\sigma'}_{\alpha\beta, \mu\nu}$ read
\begin{eqnarray}    \label{eq:53}
&& M^{\sigma\sigma'}_{\alpha\beta, \mu\nu} =  \delta_{\mu \alpha} f_{\nu \beta}^{\sigma^{\prime} *} + \delta_{\nu\beta} f_{\alpha \mu}^{\sigma^{\prime} * } ;  \nonumber \\
&& N^{\sigma\sigma'}_{\alpha\beta, \mu\nu} = \delta_{\mu \alpha} f_{\beta \nu}^{\sigma^{\prime}} + \delta_{\nu \beta} f_{\mu \alpha}^{\sigma^{\prime} } .
\end{eqnarray}
Using these definitions, we have 
\begin{equation}    \label{eq:54}
[A^{\sigma}_{\alpha \beta}, H_{1}] = \sum_{k \sigma '} \sum_{\mu\nu} V_{k\sigma '} \left[ M^{\sigma\sigma'}_{\alpha\beta, \mu\nu} A_{\mu \nu} c_{k\sigma^{\prime}}  -  N^{\sigma\sigma'}_{\alpha\beta, \mu\nu} c_{k\sigma^{\prime}}^{\dagger}A_{\mu \nu} \right].
\end{equation}
$G_{i}(A^{\sigma}_{\alpha \beta}|A_{\gamma \delta} ^{\sigma \dagger})_{\omega}$ is then obtained as 
\begin{eqnarray}    \label{eq:55}
&& G_{i}(A^{\sigma}_{\alpha \beta}|A_{\gamma \delta} ^{\sigma \dagger})_{\omega}   \nonumber \\
&=&  \frac{\delta_{\beta \delta} \langle A_{\alpha \gamma} \rangle_{i}+ \delta_{\gamma \alpha} \langle A_{\delta\beta} \rangle_{i} }{\omega + E_{\alpha} - E_{\beta}} \nonumber \\
&+& \sum_{k \sigma'}\sum_{\mu\nu} \frac{V_{k \sigma'} M^{\sigma\sigma'}_{\alpha\beta, \mu\nu} }{\omega + E_{\alpha} - E_{\beta}}G_{i-1}\left(A_{\mu\nu}c_{k \sigma'}|A_{\gamma \delta} ^{\sigma \dagger} \right)_{\omega}   \nonumber \\
&-& \sum_{k \sigma'}\sum_{\mu\nu} \frac{V_{k \sigma'} N^{\sigma\sigma'}_{\alpha\beta, \mu\nu} }{\omega + E_{\alpha} - E_{\beta}}G_{i-1}\left(c^{\dagger}_{k \sigma'}A_{\mu\nu}|A_{\gamma \delta} ^{\sigma \dagger} \right)_{\omega} .
\end{eqnarray}

For $i=1$, the above equation involves new GFs $G_{0}\left(A_{\mu \nu}c_{k \sigma'}|A_{\gamma \delta} ^{\sigma \dagger} \right)_{\omega}$ and $G_{0}\left(c^{\dagger}_{k \sigma'} A_{\mu \nu}|A_{\gamma \delta} ^{\sigma \dagger} \right)_{\omega}$. They are zero because the impurity and bath are decoupled in $H_0$. Using the sum rule $\sum_{\mu} \langle A_{\mu\mu} \rangle_{i}= \langle  1 \rangle_i = \delta_{i,0}$, the self-consistent solution gives $\langle A_{\alpha\beta} \rangle_{1} =0$. Therefore we get
\begin{equation}    \label{eq:56}
G_{1}(A^{\sigma}_{\alpha \beta}|A_{\gamma \delta} ^{\sigma \dagger})_{\omega}=0.
\end{equation}
In general, $G_{i}(A^{\sigma}_{\alpha \beta}|A_{\gamma \delta} ^{\sigma \dagger})_{\omega}=0$ for $i$ odd. This is because, starting from the impurity, an electron has to hop an even number of times to come back to the impurity and a local propagator contains only even powers of $V_k$. 

For $i=2$, Eq.(\ref{eq:55}) has the unknown averages of the type $\langle A_{\alpha \beta} \rangle_{2}$ and new first-order GFs $G_{1}(A_{\mu\nu}c_{k \sigma'}|A_{\gamma \delta} ^{\sigma \dagger} )_{\omega}  $ and $G_{1}(c^{\dagger}_{k \sigma'}A_{\mu\nu}|A_{\gamma \delta} ^{\sigma \dagger} )_{\omega}$. 
The former is to be solved self-consistently with $G_{2}(A_{\mu\nu}|A_{\gamma \delta} ^{\sigma \dagger} )_{\omega}$ later. For the latter, we can write down their EOMs and solve them similarly. Note that $A_{\mu\nu}$ here is a Grassmann-even operator. In this process, new zeroth-order GFs will be generated. They can be expressed by the known quantities $G_{0}( A_{\alpha\beta}|A_{\gamma \delta} ^{\sigma \dagger} )_{\omega} $ and $\langle c^{\dagger}_{k \sigma} c_{k \sigma} \rangle_0$, using the fact that the impurity and bath are decoupled in $H_0$. The final results are 
\begin{eqnarray}    \label{eq:57}
&&  G_{1}(A_{\mu\nu}c_{k \sigma'}|A_{\gamma \delta} ^{\sigma \dagger} )_{\omega}    \nonumber \\
&=& \frac{ -\delta_{\nu \delta} \langle A_{\mu \gamma} c_{k \sigma'}\rangle_{1} + \delta_{\mu\gamma}     
\langle A_{\delta\nu} c_{k\sigma'} \rangle_{1}  }{ \omega + E_{\mu} - E_{\nu} -\epsilon_{k\sigma'} }   \nonumber \\
&& + G_{0}(A_{\gamma \delta} | A_{\gamma \delta}^{\dagger})_{\omega} \frac{V_{k \sigma'}\left[ \delta_{\mu \gamma} f_{\nu \delta}^{\sigma'} - H_{\mu\nu, \gamma\delta}^{\sigma'} \langle n_{k\sigma'}\rangle_0  \right] }{ \omega + E_{\mu} - E_{\nu} -\epsilon_{k\sigma'} } ,    \nonumber \\
&&
\end{eqnarray}
and
\begin{eqnarray}    \label{eq:58}
&&  G_{1}( c^{\dagger}_{k \sigma'}A_{\mu\nu}|A_{\gamma \delta} ^{\sigma \dagger} )_{\omega}    \nonumber \\
&=& \frac{ \delta_{\nu \delta} \langle c^{\dagger}_{k \sigma'} A_{\mu \gamma}\rangle_{1} - \delta_{\mu\gamma}     
\langle c^{\dagger}_{k \sigma'}A_{\delta\nu} \rangle_{1} }{ \omega + E_{\mu} - E_{\nu} + \epsilon_{k\sigma'} }    \nonumber \\
&& - G_{0}( A_{\gamma \delta}|A_{\gamma \delta}^{\dagger} )_{\omega}  \frac{ V_{k \sigma'} \left[ \delta_{\nu \delta} f_{\mu \gamma}^{\sigma' *} + L_{\mu\nu, \gamma\delta}^{\sigma'}\langle n_{k\sigma'}\rangle_0 \right] }{  \omega + E_{\mu} - E_{\nu} + \epsilon_{k\sigma'} }.  \nonumber \\
&&
\end{eqnarray}
In the above equations, the newly introduced coefficients $L^{\sigma}_{\mu \nu, \lambda \tau}$ and $H^{\sigma}_{\mu \nu, \lambda \tau}$ are defined as
\begin{eqnarray}    \label{eq:59}
&& [A_{\mu \nu}, d_{\sigma}^{\dagger}] = \sum_{\lambda \tau} L^{\sigma}_{\mu\nu, \lambda \tau} A_{\lambda \tau};   \nonumber \\
&& [A_{\mu \nu}, d_{\sigma}] = \sum_{\lambda \tau} H^{\sigma}_{\mu\nu, \lambda \tau} A_{\lambda \tau}.
\end{eqnarray}
Their expressions are
\begin{eqnarray}    \label{eq:60}
&& L^{\sigma}_{\mu\nu, \lambda \tau} = \delta_{\mu \lambda} f_{\tau\nu}^{\sigma *} - \delta_{\nu \tau} f_{\mu \lambda}^{\sigma *} ;   \nonumber \\
&& H^{\sigma}_{\mu\nu, \lambda \tau} = \delta_{\mu \lambda} f_{\nu \tau}^{\sigma} - \delta_{\nu \tau} f_{\lambda \mu}^{\sigma}.
\end{eqnarray}

The four averages $\langle A_{\mu \gamma} c_{k \sigma}\rangle_{1}$, $\langle A_{\delta \nu} c_{k \sigma}\rangle_{1}$, $\langle c_{k \sigma}^{\dagger} A_{\mu \gamma}\rangle_{1}$ and $\langle c_{k \sigma}^{\dagger} A_{\delta \nu} \rangle_{1}$ in Eq.(\ref{eq:57}) and Eq.(\ref{eq:58}) need to be calculated from the GFs like $G_{1}(c_{k \sigma}|A_{\mu \gamma})_{\omega}$ etc. using the fluctuation-dissipation theorem. For $\langle A_{\mu \gamma} c_{k \sigma}\rangle_{1}$, we solve the EOM for $G_{1}(c_{k \sigma}|A_{\mu \gamma})_{\omega}$ and get
\begin{eqnarray}    \label{eq:61}
  && G_{1}(c_{k \sigma}|A_{\mu \gamma})_{\omega}  \nonumber \\
 &=& \frac{ V_{k\sigma} f_{\gamma \mu}^{\sigma}  (a_{\gamma} + a_{\mu}) }{E_{\gamma} - E_{\mu} + \epsilon_{k \sigma}} \left[\frac{1}{\omega - \epsilon_{k\sigma}} - \frac{1}{\omega + E_{\gamma} - E_{\mu}} \right],  \nonumber \\
 &&
\end{eqnarray}
which gives
\begin{eqnarray}    \label{eq:62}
 && \langle A_{\mu \gamma} c_{k \sigma}\rangle_{1}  \nonumber \\
 & =& \frac{  V_{k\sigma} f_{\gamma \mu}^{\sigma} (a_{\gamma} + a_{\mu}) }{E_{\gamma} - E_{\mu} + \epsilon_{k \sigma}} \left[\frac{1}{e^{\beta \epsilon_{k \sigma}} + 1} - \frac{1}{e^{\beta (E_{\mu} - E_{\gamma} )} +1}  \right].   \nonumber \\
&& 
\end{eqnarray}
Using $\langle c^{\dagger}_{k \sigma} A_{\mu \gamma} \rangle_{1} = \langle A_{\gamma \mu} c_{k \sigma}\rangle_{1}^{*}$ and the replacement ($\mu \rightarrow \delta, \gamma \rightarrow \nu$), the other three averages can be obtained. To carry out the $k$-summations in Eq.(\ref{eq:55}), it useful to introduce the following intermediate quantity,
\begin{eqnarray}    \label{eq:63}
   \Phi_{\alpha \beta}^{\sigma}(\omega) &\equiv& \sum_{k} \frac{V_{k\sigma} \langle A_{\alpha \beta} c_{k \sigma} \rangle_1}{\omega - \epsilon_{k \sigma}}  \nonumber \\ 
   &=& \frac{f_{\beta \alpha}^{\sigma} (a_{\alpha} + a_{\beta} )}{ \omega + E_{\beta} - E_{\alpha}} \varphi_{\beta \alpha}^{\sigma}(\omega),
\end{eqnarray}
with
\begin{equation}    \label{eq:64}
   \varphi_{\beta \alpha}^{\sigma}(\omega) = \Lambda_{\sigma}(\omega) - \Lambda_{\sigma}(E_{\alpha} - E_{\beta} ) - \frac{\Gamma_{\sigma}(\omega) - \Gamma_{\sigma}(E_{\alpha} - E_{\beta}) }{ e^{\beta(E_{\alpha} - E_{\beta})} + 1}.
\end{equation}
Here, the two involved functions are
\begin{eqnarray}    \label{eq:65}
&&   \Gamma_{\sigma}(\omega) = \int_{-\infty}^{\infty} \frac{\Delta(\epsilon)}{\omega - \epsilon} d\epsilon ;  \nonumber \\
&&    \Lambda_{\sigma}(\omega) = \int_{-\infty}^{\infty} \frac{\Delta(\epsilon)}{\omega - \epsilon} \frac{1}{e^{\beta \epsilon} + 1} d\epsilon.
\end{eqnarray}
Likewise, the Hermitian conjugate of Eq.(\ref{eq:63}) is
\begin{equation}    \label{eq:66}
  \sum_{k} \frac{V_{k\sigma} \langle c^{\dagger}_{k \sigma} A_{\alpha \beta}  \rangle_1}{\omega - \epsilon_{k \sigma}} =  \Phi_{\beta \alpha }^{\sigma}(\omega) . 
\end{equation}

With these preparations, we put Eq.(\ref{eq:57}) and (\ref{eq:58}) into Eq.(\ref{eq:55}), simplify all the terms and obtain
\begin{eqnarray}    \label{eq:67}
 &&  G_{2}(A^{\sigma}_{\alpha \beta}|A_{\gamma \delta} ^{\sigma \dagger})_{\omega}    \nonumber \\
 & = & \frac{\delta_{\beta \delta} \langle A_{\alpha \gamma} \rangle_{2}+ \delta_{\gamma \alpha} \langle A_{\delta\beta} \rangle_{2} }{\omega + E_{\alpha} - E_{\beta} }    \nonumber \\
 &+ & \frac{ J_{\alpha\beta, \gamma\delta}^{\sigma}(\omega) + F_{\alpha\beta, \gamma\delta}^{\sigma}(\omega) }{ \left( \omega + E_{\alpha} - E_{\beta} \right)\left( \omega + E_{\gamma} - E_{\delta} \right)},
\end{eqnarray}
with $J_{\alpha\beta, \gamma\delta}^{\sigma}(\omega)$ and $F_{\alpha\beta, \gamma\delta}^{\sigma}(\omega)$ given by
\begin{eqnarray}    \label{eq:68}
&&  J_{\alpha\beta, \gamma\delta}^{\sigma}(\omega)   \nonumber \\
&=& - \sum_{\mu \sigma'} M_{\alpha \beta, \mu\delta}^{\sigma \sigma'} f_{\gamma \mu}^{\sigma'} (a_{\mu} + a_{\delta}) \varphi_{\gamma \mu}^{\sigma'} (\omega + E_{\mu}- E_{\delta})    \nonumber \\
&& + \sum_{\nu \sigma'} M_{\alpha \beta, \gamma \nu}^{\sigma \sigma'} f_{\nu  \delta}^{\sigma'} (a_{\nu} + a_{\delta}) \varphi_{\nu \delta}^{\sigma'} (\omega + E_{\gamma}- E_{\nu})    \nonumber \\
&& - \sum_{\mu \sigma'} N_{\alpha \beta, \mu\delta}^{\sigma \sigma'} f_{\mu \gamma}^{\sigma'} (a_{\mu} + a_{\gamma}) \varphi_{\mu \gamma}^{\sigma'} (-\omega - E_{\mu} + E_{\delta})    \nonumber \\
&& + \sum_{\nu \sigma'} N_{\alpha \beta, \gamma \nu}^{\sigma \sigma'} f_{\delta \nu}^{\sigma'} (a_{\nu} + a_{\delta}) \varphi_{\delta \nu}^{\sigma'} (-\omega - E_{\gamma} + E_{\nu}). \nonumber \\
&&
\end{eqnarray}
\begin{eqnarray}    \label{eq:69}
&&  F_{\alpha\beta, \gamma\delta}^{\sigma}(\omega)   \nonumber \\
&=& (a_{\gamma} + a_{\delta}) \sum_{\nu \sigma'} M_{\alpha \beta, \gamma \nu}^{\sigma \sigma'} f_{\nu \delta}^{\sigma'}  \Gamma_{\sigma'} (\omega + E_{\gamma}- E_{\nu} )    \nonumber \\
&& - (a_{\gamma} + a_{\delta}) \sum_{\mu \sigma'} N_{\alpha \beta, \mu \delta}^{\sigma \sigma'} f_{\mu  \gamma}^{\sigma' *} \Gamma_{\sigma'} (-\omega - E_{\mu} + E_{\delta})    \nonumber \\
&& - (a_{\gamma} + a_{\delta}) \sum_{\mu \nu} \sum_{\sigma'} M_{\alpha \beta, \mu \nu}^{\sigma \sigma'} H_{\mu \nu, \gamma \delta}^{\sigma'}  \Lambda_{\sigma'} (\omega + E_{\mu}- E_{\nu} )    \nonumber \\
&& - (a_{\gamma} + a_{\delta}) \sum_{\mu\nu} \sum_{\sigma'} N_{\alpha \beta, \mu \nu}^{\sigma \sigma'} L_{\mu \nu, \gamma \delta}^{\sigma'}  \Lambda_{\sigma'} (-\omega - E_{\mu} + E_{\nu}). \nonumber \\
&&
\end{eqnarray}
Eqs.(\ref{eq:67})-(\ref{eq:69}) are the results for $G_{2}(A^{\sigma}_{\alpha \beta}|A_{\gamma \delta} ^{\sigma \dagger})_{\omega}$, where $ J_{\alpha\beta, \gamma\delta}^{\sigma}(\omega)$ involves $\varphi_{\alpha\beta}^{\sigma}(\omega)$ and comes from the averages $\langle c^{\dagger}_{k \sigma} A_{\alpha \beta} \rangle_1$. Note that in the calculation we have made use of the Grassmann-odd properties of $A^{\sigma}_{\alpha \beta}$. Therefore these equations do not apply to GF $G(A_{\alpha \beta}|A_{\gamma \delta}^{\dagger})_{\omega}$ with arbitrary $A_{\alpha \beta}$.

Below, we focus on the particle-hole symmetric case and simplify these equations for numerical calculation. For the single impurity Anderson model considered in this paper, the four eigen states of the local impurity Hamiltonian $\hat{h}_0 \equiv U n_{\uparrow} n_{\downarrow} - \mu \sum_{\sigma} n_{\sigma}$ are $|1 \rangle = d_{\uparrow}^{\dagger}| 0\rangle$, $|2\rangle = d_{\downarrow}^{\dagger}| 0\rangle$, $|3\rangle =|0\rangle$ , and  $|4 \rangle = d_{\uparrow}^{\dagger}d_{\downarrow}^{\dagger}| 0\rangle$. The corresponding eigen energies are $E_{1} = E_{2} = \epsilon_d$, $E_{3}=0$, and $E_{4}=U+2\epsilon_d$. There are total $16$ SBOs $A_{\alpha \beta}$ ($\alpha, \beta = 1 \sim 4$). The impurity electron annihilation operator is expanded as $d_{\uparrow} = A_{31} + A_{24}$ and $d_{\downarrow} = A_{32} - A_{14}$, meaning $f^{\uparrow}_{31} = f^{\uparrow}_{24}=1$ and $f^{\downarrow}_{32} = -f^{\downarrow}_{14} = 1$ and others are zero. In terms of electron operators, $A_{11} = n_{\uparrow}(1-n_{\downarrow})$, $A_{22} = (1-n_{\uparrow})n_{\downarrow}$, $A_{33} = (1-n_{\uparrow})(1-n_{\downarrow})$, and $A_{44} =  n_{\downarrow}n_{\uparrow}$. Their averages play important role in the $T$-dependence of GFs.

The particle-hole symmetry implies the following conditions
\begin{eqnarray}    \label{eq:70}
 && \epsilon_d= -U/2;     \nonumber \\
 && \Delta_{\sigma}(-\omega) = \Delta_{\bar{\sigma}}(\omega).
\end{eqnarray} 
We have $E_{1}=E_{2}=-U/2$ and $E_{3}=E_{4}=0$ to simplify Eqs.(\ref{eq:50}) and (\ref{eq:67})-(\ref{eq:69}). The local GF up to order $V_{k}^2$ reads $G(d_{\sigma}|d_{\sigma}^{\dagger})_{\omega}  \approx G_{0}(d_{\sigma}|d_{\sigma}^{\dagger})_{\omega}  + G_{2}(d_{\sigma}|d_{\sigma}^{\dagger})_{\omega}$, with
\begin{eqnarray}    \label{eq:71}
G_{0}(d_{\sigma}|d_{\sigma}^{\dagger})_{\omega} &=& \frac{1/2}{\omega + U/2} + \frac{1/2}{\omega - U/2};  \nonumber \\
G_{2}(d_{\sigma}|d_{\sigma}^{\dagger})_{\omega}  &=& \frac{W_{1}^{\sigma}-1/2}{\omega + U/2} + \frac{W_{2}^{\sigma}-1/2}{\omega - U/2}  + \frac{W_{3}^{\sigma}(\omega)}{ (\omega + U/2)^2 }  \nonumber \\
  && +\frac{W_{4}^{\sigma}(\omega)}{ (\omega - U/2)^2 }  + \frac{W_{5}^{\sigma}(\omega)}{(\omega + U/2)(\omega - U/2) }.  \nonumber \\
&&  
\end{eqnarray}
The weights $W_{i}^{\sigma}$ ($i=1 \sim 5$) for spin up are
\begin{eqnarray}    \label{eq:72}
   W_{1}^{\uparrow} &=& \frac{1}{2} + \langle A_{11} \rangle_{2} + \langle A_{33} \rangle_{2};   \nonumber \\
   W_{2}^{\uparrow} &=& \frac{1}{2} + \langle A_{22} \rangle_{2} + \langle A_{44} \rangle_{2};   \nonumber \\
   W_{3}^{\uparrow}(\omega) &=& \frac{1}{2}\left[\varphi_{14}^{\downarrow}(-\omega) - \varphi_{32}^{\downarrow}(\omega) \right]   \nonumber \\
 && + \frac{1}{2} \left[ \Gamma_{\uparrow}(\omega) +  \Lambda_{\downarrow}(\omega) - \Lambda_{\downarrow}(-\omega) \right];   \nonumber \\
 W_{4}^{\uparrow}(\omega) &=& \frac{1}{2}\left[\varphi_{14}^{\downarrow}(\omega) - \varphi_{32}^{\downarrow}(-\omega) \right]   \nonumber \\
 && + \frac{1}{2} \left[ \Gamma_{\uparrow}(\omega) + \Gamma_{\downarrow}(\omega) - \Gamma_{\downarrow}(-\omega) - \Lambda_{\downarrow}(\omega) +  \Lambda_{\downarrow}(-\omega) \right]; \nonumber \\
  W_{5}^{\uparrow}(\omega) &=& \frac{1}{2}\left[-\varphi_{14}^{\downarrow}(\omega) + \varphi_{32}^{\downarrow}(-\omega)-\varphi_{14}^{\downarrow}(-\omega) + \varphi_{32}^{\downarrow}(\omega) \right]   \nonumber \\
 && + \frac{1}{2} \left[ -\Gamma_{\downarrow}(\omega) + \Gamma_{\downarrow}(-\omega) \right].    
\end{eqnarray}
For spin down, 
\begin{eqnarray}    \label{eq:73}
   W_{1}^{\downarrow} &=& \frac{1}{2} + \langle A_{22} \rangle_{2} + \langle A_{33} \rangle_{2};   \nonumber \\
   W_{2}^{\downarrow} &=& \frac{1}{2} + \langle A_{11} \rangle_{2} + \langle A_{44} \rangle_{2};   \nonumber \\
   W_{3}^{\downarrow}(\omega) &=& \frac{1}{2}\left[\varphi_{24}^{\uparrow}(-\omega) - \varphi_{31}^{\uparrow}(\omega) \right]   \nonumber \\
 && + \frac{1}{2} \left[ \Gamma_{\downarrow}(\omega) +  \Lambda_{\uparrow}(\omega) - \Lambda_{\uparrow}(-\omega) \right];   \nonumber \\
 W_{4}^{\downarrow}(\omega) &=& \frac{1}{2}\left[\varphi_{24}^{\uparrow}(\omega) - \varphi_{31}^{\uparrow}(-\omega) \right]   \nonumber \\
 && + \frac{1}{2} \left[ \Gamma_{\downarrow}(\omega) + \Gamma_{\uparrow}(\omega) - \Gamma_{\uparrow}(-\omega) - \Lambda_{\uparrow}(\omega) +  \Lambda_{\uparrow}(-\omega) \right]; \nonumber \\
  W_{5}^{\uparrow}(\omega) &=& -\frac{1}{2}\left[\varphi_{24}^{\uparrow}(\omega) - \varphi_{31}^{\uparrow}(-\omega) + \varphi_{24}^{\uparrow}(-\omega) - \varphi_{31}^{\uparrow}(\omega) \right]   \nonumber \\
 && - \frac{1}{2} \left[ \Gamma_{\uparrow}(\omega) - \Gamma_{\uparrow}(-\omega) \right].  
\end{eqnarray}
Note that only the averages of diagonal SBOs $\langle A_{\alpha \alpha} \rangle$ are involved in $G(d_{\sigma}|d_{\sigma}^{\dagger})_{\omega}$. This is due to the fact that both $d_{\uparrow} = A_{31} + A_{24}$ and $d_{\downarrow} = A_{32} - A_{14}$ consist of SBOs with nonoverlap subscripts. It is found that the relation $W_{3}^{\sigma} + W_{4}^{\sigma} + W_{5}^{\sigma} = \Gamma_{\sigma}(\omega)$ holds here. The second condition in Eq.(\ref{eq:70}) implies $\Gamma_{\sigma}(-\omega) = - \Gamma_{\bar{\sigma}}(\omega)$ and $\Lambda_{\sigma}(-\omega) = \Lambda_{\bar{\sigma}}(\omega) - \Gamma_{\bar{\sigma}}(\omega)$ which can further simplify the expressions for $W_{i}^{\sigma}(\omega)$ ($i=3 \sim 5$). Especially, we find $W_{3}^{\sigma}(\omega) = W_{4}^{\sigma}(\omega)$.

Now let us consider the self-consistent determination of the averages $\langle A_{\alpha\alpha} \rangle_{2}$ ($\alpha=1 \sim 4$) which appear in $W_{1}^{\sigma}$ and $W_{2}^{\sigma}$. They can be calculated from $G_{2}(A_{31}|A_{31}^{\dagger})_{\omega}$, $G_{2}(A_{32}|A_{32}^{\dagger})_{\omega}$, $G_{2}(A_{24}|A_{24}^{\dagger})_{\omega}$, and $G_{2}(A_{14}|A_{14}^{\dagger})_{\omega}$, respectively. Using $W_{i}^{\sigma}$ ($i=1 \sim 5$) defined above, these GFs read
\begin{eqnarray}    \label{eq:74}
&&  G_{2}(A_{31}|A_{31}^{\dagger})_{\omega} = \frac{W_{1}^{\uparrow} - 1/2}{\omega + U/2} + \frac{W_{3}^{\uparrow}(\omega)}{ (\omega + U/2)^2 }  ; \nonumber \\
&&  G_{2}(A_{24}|A_{24}^{\dagger})_{\omega} = \frac{W_{2}^{\uparrow} - 1/2}{\omega - U/2} + \frac{W_{4}^{\uparrow}(\omega)}{ (\omega - U/2)^2 }  ; \nonumber \\
&& G_{2}(A_{32}|A_{32}^{\dagger})_{\omega} = \frac{W_{1}^{\downarrow} - 1/2}{\omega + U/2} + \frac{W_{3}^{\downarrow}(\omega)}{ (\omega + U/2)^2 }  ; \nonumber \\
&& G_{2}(A_{14}|A_{14}^{\dagger})_{\omega} = \frac{W_{2}^{\downarrow} - 1/2}{\omega - U/2} + \frac{W_{4}^{\downarrow}(\omega)}{ (\omega - U/2)^2 }  .
\end{eqnarray}
From these GFs, the self-consistent equations for $\langle A_{\alpha\alpha} \rangle_{2}$ ($\alpha=1 \sim 4$) are obtained as
\begin{eqnarray}    \label{eq:75}
&& \frac{\langle A_{11} \rangle_{2}}{e^{\beta U/2}+1} - \frac{ \langle A_{33} \rangle_{2}}{e^{-\beta U/2} + 1} = \left\langle \frac{W_{3}^{\uparrow}(\omega)}{ (\omega + U/2)^2 } \right\rangle;  \nonumber \\   
&& \frac{\langle A_{44} \rangle_{2}}{e^{-\beta U/2}+1} - \frac{ \langle A_{22} \rangle_{2}}{e^{\beta U/2} + 1} = \left\langle \frac{W_{4}^{\uparrow}(\omega)}{ (\omega - U/2)^2 } \right\rangle; \nonumber \\   
&& \frac{\langle A_{22} \rangle_{2}}{e^{\beta U/2}+1} - \frac{ \langle A_{33} \rangle_{2}}{e^{-\beta U/2} + 1} = \left\langle \frac{W_{3}^{\downarrow}(\omega)}{ (\omega + U/2)^2 } \right\rangle; \nonumber \\   
&& \frac{\langle A_{44} \rangle_{2}}{e^{-\beta U/2}+1} - \frac{ \langle A_{11} \rangle_{2}}{e^{\beta U/2} + 1} = \left\langle \frac{W_{4}^{\downarrow}(\omega)}{ (\omega - U/2)^2 } \right\rangle.
\end{eqnarray}
In the above equations, the symbol $\langle g(\omega)\rangle$ is defined as
$\langle g(\omega) \rangle \equiv -1/\pi \int_{-\infty}^{\infty} {\text Im} g(\omega + i\eta) 1/(e^{\beta \omega} + 1) d\omega$. Note that the four equations in Eq.(\ref{eq:75}) are not independent. One needs to supplement the forth independent equation, i.e.,
\begin{equation}    \label{eq:76}
   \langle A_{11} \rangle_2 +    \langle A_{22} \rangle_2 +   \langle A_{33} \rangle_2 +   \langle A_{44} \rangle_2  = \langle 1 \rangle_2 = 0.
\end{equation}
These equations have to be solved numerically except for the case of paramagnetic bath where analytical solution is given below.
In the limit $T=0$, care must be taken in evaluating the right-hand side of Eq.(\ref{eq:77}) and solving the equations, because the diverging $\beta$ factors in the averages $\langle A_{\alpha\alpha} \rangle_2$ ($\alpha=1 \sim 4$) can make the numerical process unstable. In our numerical calculation, we isolate the most singular item from the above equations to stabilize the solution.

In the paramagnetic phase, $\Delta_{\uparrow}(\omega) = \Delta_{\downarrow}(\omega)$. One finds $\varphi_{14}^{\downarrow}(-\omega) = \varphi_{32}^{\downarrow}(\omega)$ and $\varphi_{24}^{\uparrow}(-\omega) = \varphi_{31}^{\uparrow}(\omega)$. Eqs.(\ref{eq:72}) and (\ref{eq:73}) give
$W_{1}^{\sigma} = W_{2}^{\sigma} = 1/2$, $W_{3}^{\sigma}(\omega) = W_{4}^{\sigma}(\omega) = \Gamma(\omega)$, and $W_{5}^{\sigma}(\omega) = - \Gamma(\omega)$. The self-consistent calculation of $A_{\alpha\alpha}$ in Eq.(\ref{eq:75}) is no longer necessary. We obtain $G(d_{\sigma}|d_{\sigma}^{\dagger})_{\omega}  = G_{0}(d_{\sigma}|d_{\sigma}^{\dagger})_{\omega}  + G_{2}(d_{\sigma}|d_{\sigma}^{\dagger})_{\omega} + ...$ and
\begin{eqnarray}    \label{eq:77}
   G_{0}(d_{\sigma}|d_{\sigma}^{\dagger})_{\omega}  &=& \frac{1/2}{\omega + U/2} + \frac{1/2}{\omega - U/2};  \nonumber \\
  G_{2}(d_{\sigma}|d_{\sigma}^{\dagger})_{\omega}  &=&     \frac{\Gamma(\omega)}{ (\omega + U/2)^2 } +\frac{\Gamma(\omega)}{ (\omega - U/2)^2 }  \nonumber \\
  && - \frac{\Gamma(\omega)}{(\omega + U/2)(\omega - U/2) }.  \nonumber \\
&&  
\end{eqnarray}
This recovers the GF obtained by direct expansion in Ref.~\onlinecite{Dai1}.

\subsection{SE-resummation and CF-resummation}
In this subsection, we do the resummation for the second-order strong-coupling expansion obtained above. The most frequently used resummation method is via SE Eq.(\ref{eq:19}). Although we have argued that SE resummation cannot solve the causality or the zero temperature divergence problem in general, here we will show the data for comparison. The second method that we will use is the CF resummation method Eq.(\ref{eq:20}) which is guaranteed to be causal but may have the zero temperature divergence problem.

For SE resummation, from Eq.(\ref{eq:71}) we obtain up to order $V_{k}^{2}$ 
\begin{eqnarray}   \label{eq:78}
&&   \Sigma_{0\sigma}(\omega) = U/2 + (U/2)^2 /\omega; \nonumber \\
&&  \Sigma_{1\sigma}(\omega) = 0;     \nonumber \\
&&  \Sigma_{2\sigma}(\omega) = G_{0}^{-2}(d_{\sigma}|d_{\sigma}^{\dagger})_{\omega}
G_{2}(d_{\sigma}|d_{\sigma}^{\dagger})_{\omega} - \Gamma_{\sigma}(\omega).
\end{eqnarray}
The SE-resummed GF is then obtained by inserting this SE into the Dyson equation $\overline{G}^{SE}(d_{\sigma}|d_{\sigma}^{\dagger})_{\omega} = \left[ \mathcal{G}_{0 \sigma}^{-1}(\omega) - \Sigma_{\sigma}(\omega) \right]^{-1}$. One gets\cite{Dai1}
\begin{eqnarray}   \label{eq:79}
&& \overline{G}_{SE}(d_{\sigma}|d_{\sigma}^{\dagger})_{\omega} \nonumber \\
&=& \frac{1}{\omega +U/2 - \Gamma_{\sigma}(\omega) - \Sigma_{0\sigma}(\omega)- \Sigma_{2 \sigma}(\omega) }.
\end{eqnarray}

To do the CF resummation, we first expand Eq.(\ref{eq:71}) into Taylor series of $1/\omega$. Note that the $\omega$-dependence in $W_{i}^{\sigma}(\omega)$ ($i=3,4,5$) arises solely from the hybridization function $\Delta(\epsilon)$ which is proportional to $V_{k}^2$. We therefore only expand $\omega$ in the denominators of Eq.(\ref{eq:71}) and treat $W_{i}^{\sigma}(\omega)$ $(i=3,4,5)$ as constants. The obtained series is compared with the same expansion of $\overline{G}_{CF}(d_{\sigma}|d_{\sigma}^{\dagger})_{\omega}$ in Eq.(\ref{eq:20}). By requiring that for every $n \in [1, \infty]$, $\omega^{-n}$ terms in the two GFs agree on the level of $V_{k}^2$, we obtain the expansion of the coefficients $a_{0}, a_{1}, ...$ and $b_{1}, b_{2}, ...$. It turns out that only two levels of fraction are sufficient to match the GFs up to $V_{k}^2$. We get
\begin{equation}    \label{eq:80}
 \overline{G}_{CF}(d_{\sigma}|d_{\sigma}^{\dagger})_{\omega}  
 = \cfrac{a_0}{\omega + b_1
          - \cfrac{a_1}{\omega + b_2 } }, 
\end{equation}
with coefficients
\begin{eqnarray}   \label{eq:81}
&&  a_{0} = 1;  \nonumber \\
&&  a_{1} = \left(U/2 \right)^2;  \nonumber \\
&&  b_{1} = \frac{U}{2} (W_{1}^{\sigma} - W_{2}^{\sigma}) - \Gamma_{\sigma}(\omega);   \nonumber \\
&&  b_{2} = -\frac{U}{2}(W_{1}^{\sigma} - W_{2}^{\sigma}) - \Gamma_{\sigma}(\omega) + 2W_{5}^{\sigma}(\omega).
\end{eqnarray}
Here, $a_{0} \geq 0$ and $a_{1} \geq 0$ guarantees that $\overline{G}_{CF}(d_{\sigma}|d_{\sigma}^{\dagger})_{\omega}$ has real simple poles and is causal. If we also expand $\omega$ in $W_{i}^{\sigma}$ ($i=3,4,5$) to make the comparison, due to the continuous bath degrees of freedom, a CF with infinite number of levels will be required to match Eq.(\ref{eq:71}) to $V_{k}^{2}$ order.

$W_{1}^{\sigma}$ and $W_{2}^{\sigma}$ enters the pole position of $\overline{G}_{SE}(d_{\sigma}|d_{\sigma}^{\dagger})_{\omega}  $ and $\overline{G}_{CF}(d_{\sigma}|d_{\sigma}^{\dagger})_{\omega}$, either through $\Sigma_{2\sigma}(\omega)$ in Eq.(\ref{eq:79}) or through $b_1$ and $b_2$ in Eq.(\ref{eq:80}). Since they contain a $\beta$ factor, both GFs have the zero-temperature divergence problem as shown below.

For the paramagnetic bath, SE-resummation and CF-resummation give respectively 
\begin{equation}   \label{eq:82}
   \overline{G}_{SE}(d_{\sigma}|d_{\sigma}^{\dagger})_{\omega} = \frac{1}{\omega - \Gamma(\omega) - \left(\frac{U}{2}\right)^2  \left[\frac{1}{\omega} + \frac{3\Gamma(\omega)}{\omega^2} \right] },
\end{equation}
and 
\begin{equation}   \label{eq:83}
   \overline{G}_{CF}(d_{\sigma}|d_{\sigma}^{\dagger})_{\omega} = \frac{1}{\omega - \Gamma(\omega) - \left(\frac{U}{2}\right)^2/ \left[ \omega - 3\Gamma(\omega) \right] }.
\end{equation}
Eq.(\ref{eq:82}) was obtained by a direct expansion method in Ref.~\onlinecite{Dai1} and Eq.(\ref{eq:83}) was proposed there as an {\it ad hoc} remedy of the causality problem. It is seen that $\overline{G}_{SE}(d_{\sigma}|d_{\sigma}^{\dagger})_{\omega}$ has a complex pole, leading to violation of sum rule in the local spectral function $ \rho_{\sigma}^{SE}(\omega)$, as shown below. $\overline{G}_{CF}(d_{\sigma}|d_{\sigma}^{\dagger})_{\omega}$ has real simple poles only and it conserves the rum rule. Because of $W_{1}^{\sigma}=W_{2}^{\sigma}=1/2$ in the paramagnetic bath, in both results, the zero temperature divergence problem does not appear.

\subsection{Self-Consistent Strong-Coupling Expansion to $V^2_k$ Order}

In this subsection, to remove the zero-temperature divergence problem in the bare expansion both with SE and CF resummation, we calculate $G(d_{\sigma}|d_{\sigma}^{\dagger})_{\omega}$ to order $V_{k}^2$ using the self-consistent EOM expansion method. 

We split the Anderson impurity model $H_{Aim} = H_{0} + H_{1}$ as in the bare strong-coupling expansion and use the same SBO definition. In the following, we use the self-consistent EOM expansion method described by Eq.(\ref{eq:21}) and (\ref{eq:24}). The commutators and anti-commutators involved in the calculation are same as in Sec.IV.A. We will skip the calculation details whenever they are the same as before. 

The zeroth-order GF can be obtained easily from its EOM and we obtain
\begin{equation}   \label{eq:84}
  G_{0}(A^{\sigma}_{\alpha \beta}|A_{\gamma \delta}^{\sigma \dagger})_{\omega} = \frac{\delta_{\alpha\gamma} \langle A_{\delta \beta} \rangle + \delta_{\beta \delta} \langle A_{\alpha \gamma} \rangle}{\omega + E_{\alpha} -E_{\beta}}.
\end{equation}
Here $\langle A_{\alpha \beta} \rangle$ is with respect to the full Hamiltonian $H_{Aim}$. 

The first-order renormalized contribution is similar to Eq.(\ref{eq:55}) with $i=1$ but without the average terms,
\begin{eqnarray}   \label{eq:85}
&& G_{1}(A^{\sigma}_{\alpha \beta}|A_{\gamma \delta} ^{\sigma \dagger})_{\omega}   \nonumber \\
&=&  \sum_{k \sigma'}\sum_{\mu\nu} \frac{V_{k \sigma'} M^{\sigma\sigma'}_{\alpha\beta, \mu\nu} }{\omega + E_{\alpha} - E_{\beta}}G_{0}\left(A_{\mu\nu}c_{k \sigma'}|A_{\gamma \delta} ^{\sigma \dagger} \right)_{\omega}   \nonumber \\
&-& \sum_{k \sigma'}\sum_{\mu\nu} \frac{V_{k \sigma'} N^{\sigma\sigma'}_{\alpha\beta, \mu\nu} }{\omega + E_{\alpha} - E_{\beta}}G_{0}\left(c^{\dagger}_{k \sigma'}A_{\mu\nu}|A_{\gamma \delta} ^{\sigma \dagger} \right)_{\omega} .
\end{eqnarray}
The new zeroth-order GFs $G_{0}\left(A_{\mu\nu}c_{k \sigma'}|A_{\gamma \delta} ^{\sigma \dagger} \right)_{\omega}$ and $G_{0}\left(c^{\dagger}_{k \sigma'}A_{\mu\nu}|A_{\gamma \delta} ^{\sigma \dagger} \right)_{\omega}$ can be solved from their EOM as
\begin{equation}   \label{eq:86}
   G_{0}\left(A_{\mu\nu}c_{k \sigma'}|A_{\gamma \delta} ^{\sigma \dagger} \right)_{\omega} = \frac{-\delta_{\nu\delta} \langle A_{\gamma\mu}^{\sigma \dagger} c_{k \sigma'} \rangle + \delta_{\gamma \mu}\langle  A_{\nu\delta}^{\sigma \dagger} c_{k \sigma'} \rangle }{\omega - \epsilon_{k\sigma'} + E_{\mu} - E_{\nu}},
\end{equation}
and
\begin{equation}   \label{eq:87}
   G_{0}\left(c_{k \sigma'}^{\dagger} A_{\mu\nu}|A_{\gamma \delta} ^{\sigma \dagger} \right)_{\omega} = \frac{ \delta_{\nu\delta} \langle c_{k \sigma'}^{\dagger} A_{\mu \gamma}^{\sigma} \rangle - \delta_{\gamma \mu}\langle  c_{k \sigma'}^{\dagger} A_{\delta \nu }^{\sigma} \rangle }{\omega + \epsilon_{k\sigma'} + E_{\mu} - E_{\nu}}.
\end{equation}
Differing from the bare EOM expansion method, here the averages like $\langle A_{\gamma\mu}^{\sigma \dagger} c_{k \sigma'} \rangle$ are with respect to the full Hamiltonian $H_{Aim}$ and non-zero in general. 
Putting Eqs.(\ref{eq:86})-(\ref{eq:87}) into Eq.(\ref{eq:85}), we obtain the renormalized first order GF as
\begin{eqnarray}   \label{eq:88}
   && G_{1}(A^{\sigma}_{\alpha \beta}|A_{\gamma \delta} ^{\sigma \dagger})_{\omega}   \nonumber \\
&=&   \sum_{k \sigma'}\sum_{\mu\nu} \frac{V_{k \sigma'} M^{\sigma\sigma'}_{\alpha\beta, \mu\nu} \left[-\delta_{\nu\delta} \langle A_{\gamma\mu}^{\sigma \dagger} c_{k \sigma'} \rangle + \delta_{\gamma \mu}\langle  A_{\nu\delta}^{\sigma \dagger} c_{k \sigma'} \rangle \right] }{(\omega + E_{\alpha} - E_{\beta})(\omega - \epsilon_{k\sigma'} + E_{\mu} - E_{\nu})}  \nonumber \\
&& -   \sum_{k \sigma'}\sum_{\mu\nu} \frac{V_{k \sigma'} N^{\sigma\sigma'}_{\alpha\beta, \mu\nu}  \left[\delta_{\nu\delta} \langle c_{k \sigma'}^{\dagger} A_{\mu \gamma}^{\sigma} \rangle - \delta_{\gamma \mu}\langle  c_{k \sigma'}^{\dagger} A_{\delta \nu }^{\sigma} \rangle \right] }{(\omega + E_{\alpha} - E_{\beta})(\omega + \epsilon_{k\sigma'} + E_{\mu} - E_{\nu})}. \nonumber \\
&&
\end{eqnarray}
It contains contributions in all orders of $V_{k\sigma}$ through the averages like $\langle A_{\gamma\mu}^{\sigma \dagger} c_{k \sigma'} \rangle$.

The self-consistent EOM for the second order GF is solved in a similar way as the bare one and we obtain
\begin{eqnarray}  \label{eq:89}
   && G_{2}(A^{\sigma}_{\alpha \beta}|A_{\gamma \delta} ^{\sigma \dagger})_{\omega}   \nonumber \\
&=&   \sum_{k \sigma'}\sum_{\mu\nu} \frac{V_{k \sigma'} M^{\sigma\sigma'}_{\alpha\beta, \mu\nu} G_{0}\left( [A_{\mu\nu} c_{k\sigma'}, H_{1} ]|A_{\gamma \delta} ^{\sigma \dagger} \right)_{\omega} }{(\omega + E_{\alpha} - E_{\beta})(\omega - \epsilon_{k\sigma'} + E_{\mu} - E_{\nu})}  \nonumber \\
&& -   \sum_{k \sigma'}\sum_{\mu\nu} \frac{V_{k \sigma'} N^{\sigma\sigma'}_{\alpha\beta, \mu\nu}  G_{0} \left( [c_{k\sigma'}^{\dagger} A_{\mu\nu}, H_{1} ] |A_{\gamma \delta} ^{\sigma \dagger} \right)_{\omega} }{(\omega + E_{\alpha} - E_{\beta})(\omega + \epsilon_{k\sigma'} + E_{\mu} - E_{\nu})}. \nonumber \\
&&
\end{eqnarray}
Two new GFs of zeroth order appear and their EOM can be solved to give
\begin{eqnarray}  \label{eq:90}
&&  G_{0}\left( [A_{\mu\nu} c_{k\sigma'}, H_{1} ]|A_{\gamma \delta} ^{\sigma \dagger} \right)_{\omega}  \nonumber \\
&=& V_{k\sigma'} \sum_{\tau} f_{\nu \tau}^{\sigma'} \frac{ \delta_{\tau \delta} \langle A_{\mu \gamma} \rangle + \delta_{\mu \gamma} \langle A_{\delta \tau} \rangle}{\omega + E_{\mu} - E_{\tau}}   \nonumber \\
&& + \sum_{p \sigma''}\sum_{\lambda \tau} \frac{V_{p\sigma''} L_{\mu\nu, \lambda \tau}^{\sigma''} \left\langle \left( \delta_{\tau \delta}  A_{\lambda \gamma} + \delta_{\lambda \gamma} A_{\delta \tau} \right) c_{p \sigma''} c_{k \sigma'} \right\rangle  }{ \omega + E_{\lambda} - E_{\tau} - \epsilon_{p\sigma''} - \epsilon_{k \sigma'} }  \nonumber \\
&& - \sum_{p \sigma''}\sum_{\lambda \tau} \frac{V_{p\sigma''} H_{\mu\nu, \lambda \tau}^{\sigma''} \left\langle \left( \delta_{\tau \delta}  A_{\lambda \gamma} + \delta_{\lambda \gamma} A_{\delta \tau} \right)  c_{p \sigma''}^{\dagger} c_{k \sigma'}  \right\rangle  }{ \omega + E_{\lambda} - E_{\tau} + \epsilon_{p\sigma''} - \epsilon_{k \sigma'} },   \nonumber  \\
&&
\end{eqnarray}
and
\begin{eqnarray}  \label{eq:91}
&&  G_{0}\left( [c_{k\sigma'}^{\dagger} A_{\mu\nu} , H_{1} ]|A_{\gamma \delta} ^{\sigma \dagger} \right)_{\omega}  \nonumber \\
&=& - V_{k\sigma'} \sum_{\tau} f_{\mu \tau}^{\sigma' * } \frac{ \delta_{\nu \delta} \langle A_{\tau \gamma} \rangle + \delta_{\gamma \tau} \langle A_{\delta \nu} \rangle}{\omega + E_{\tau} - E_{\nu}}   \nonumber \\
&& - \sum_{p \sigma''}\sum_{\lambda \tau} \frac{V_{p\sigma''} L_{\mu\nu, \lambda \tau}^{\sigma''} \left\langle c_{k \sigma'}^{\dagger} c_{p \sigma''} \left( \delta_{\tau \delta}  A_{\lambda \gamma} + \delta_{\lambda \gamma} A_{\delta \tau} \right)  \right\rangle  }{ \omega + E_{\lambda} - E_{\tau} + \epsilon_{k\sigma'} - \epsilon_{p \sigma''} }  \nonumber \\
&& + \sum_{p \sigma''}\sum_{\lambda \tau} \frac{V_{p\sigma''} H_{\mu\nu, \lambda \tau}^{\sigma''} \left\langle c_{k \sigma'}^{\dagger} c_{p \sigma''}^{\dagger} \left( \delta_{\tau \delta}  A_{\lambda \gamma} + \delta_{\lambda \gamma} A_{\delta \tau} \right)  \right\rangle  }{ \omega + E_{\lambda} - E_{\tau} + \epsilon_{k \sigma'} + \epsilon_{p \sigma''} }.   \nonumber  \\
&&
\end{eqnarray}

To complete the full self-consistent EOM series expansion, the averages on the right-hand sides of Eqs.(\ref{eq:84}), (\ref{eq:88}), (\ref{eq:90}) and (\ref{eq:91}) are to be calculated from the corresponding full GFs, which themselves should be obtained from a finite order EOM expansion and CF resummation. In this way, each average is obtained from a causal GF and contains $\beta$ only on the exponent. This process has certain degrees of variance because one can choose the expansion order for the GFs used to calculate these averages. Given that the target GF $G(A^{\sigma}_{\alpha \beta}|A_{\gamma \delta}^{\sigma \dagger})_{\omega}$ is produced rigorously up to $V_{k}^2$, different ways of calculating the averages amount to different approximations for the higher order contributions.

Here, to avoid further complication, we will calculate the averages in the simplest way that keeps $G(A^{\sigma}_{\alpha \beta}|A_{\gamma \delta}^{\sigma \dagger})_{\omega}$ exact up to $V_{k}^{2}$.
For this purpose, the averages $\langle A_{\delta \beta} \rangle$ and $\langle A_{\alpha \gamma} \rangle$ in Eq.(\ref{eq:84}) should be calculated from the CF-resummed GF $\overline{G}_{CF}(A^{\sigma}_{\alpha \beta}|A_{\gamma \delta}^{\sigma \dagger})_{\omega}$.
Since the averages in $G_{1}(A^{\sigma}_{\alpha \beta}|A_{\gamma \delta}^{\sigma \dagger})_{\omega}$ of Eq.(\ref{eq:88}) already has a factor $V_k$ in front of them, they will be calculated accurately to the $V_k$ level. For example, $\langle A_{\gamma\mu}^{\sigma \dagger} c_{k \sigma'} \rangle$ can be calculated from the approximate GF $G(c_{k \sigma'}|A_{\gamma\mu}^{\sigma \dagger})_{\omega} \approx G_{0}(c_{k \sigma'}|A_{\gamma\mu}^{\sigma \dagger})_{\omega} + G_{1}(c_{k \sigma'}|A_{\gamma\mu}^{\sigma \dagger})_{\omega}$. Using the self-consistent EOM expansion, we obtain
\begin{eqnarray}  \label{eq:92}
  G_{0}(c_{k \sigma'}|A_{\gamma\mu}^{\sigma \dagger})_{\omega} &=& 0; \nonumber \\
  G_{1}(c_{k \sigma'}|A_{\gamma\mu}^{\sigma \dagger})_{\omega} &=&  \frac{ V_{k\sigma'}}{\omega - \epsilon_{k \sigma'}} \sum_{\alpha \beta} f_{\alpha \beta}^{\sigma'} G_{0}(A_{\alpha \beta}^{\sigma'} | A_{\gamma\mu}^{\sigma \dagger})_{\omega}.   \nonumber \\
  &&
\end{eqnarray}
In the second equation above, $ G_{0}(A_{\alpha \beta}^{\sigma'} | A_{\gamma\mu}^{\sigma \dagger})_{\omega}$ is the renormalized zeroth-order contribution Eq.(\ref{eq:84}). Note that these contributions have real simple poles already and the CF-resummation is not necessary. After $\langle A_{\gamma\mu}^{\sigma \dagger} c_{k \sigma'} \rangle$ is obtained from the above GF, other averages in Eq.(\ref{eq:88}) can be obtained by subscript exchange, such as $\langle c_{k \sigma'}^{\dagger} A_{\mu \gamma}^{\sigma} \rangle = \langle A_{\gamma \mu}^{\sigma } c_{k \sigma'} \rangle^{*}$. 

The averages in $G_{2}(A^{\sigma}_{\alpha \beta}|A_{\gamma \delta}^{\sigma \dagger})_{\omega}$ of Eqs.(\ref{eq:90}) and (\ref{eq:91}) appear on the level of $V_{k}^2$. We will use a truncation approximation which is valid at the order $V_{k}^0$,
\begin{eqnarray}  \label{eq:93}
&& \langle A_{\lambda \gamma} c_{p \sigma''} c_{k \sigma'} \rangle \approx \langle A_{\lambda \gamma} \rangle \langle  c_{p \sigma''} c_{k \sigma'} \rangle_{0} = 0;    \nonumber \\
&& \langle c_{p \sigma''}^{\dagger} c_{k \sigma'} A_{\lambda \gamma} \rangle \approx \langle c_{p \sigma''}^{\dagger} c_{k \sigma'} \rangle_{0} \langle A_{\lambda \gamma} \rangle.
\end{eqnarray}
Similar decoupling approximations are used for other averages in Eq.(\ref{eq:90}) and those in Eq.(\ref{eq:91}). Putting these approximations into Eqs.(\ref{eq:90}), (\ref{eq:91}), and Eq.(\ref{eq:89}), the second-order GF is obtained as
\begin{eqnarray}  \label{eq:94}
   && G_{2}(A^{\sigma}_{\alpha \beta}|A_{\gamma \delta} ^{\sigma \dagger})_{\omega}   \nonumber \\
   &=& \sum_{\mu\nu}\sum_{\tau\sigma'} \frac{M_{\alpha\beta, \mu\nu}^{\sigma \sigma'} f_{\nu\tau}^{\sigma'} \Gamma_{\sigma'}(\omega + E_{\mu}-E_{\nu}) }{ \omega +E_{\alpha} - E_{\beta}}  G_{0}(A_{\mu \tau}|A_{\gamma \delta} ^{\sigma \dagger})_{\omega}  \nonumber \\
   &-& \sum_{\mu\nu}\sum_{\lambda \tau \sigma'} \frac{M_{\alpha\beta, \mu\nu}^{\sigma \sigma'} H_{\mu \nu, \lambda \tau}^{\sigma'} \Lambda_{\sigma'}(\omega + E_{\mu}-E_{\nu}) }{ \omega +E_{\alpha} - E_{\beta}}  G_{0}(A_{\lambda \tau}|A_{\gamma \delta} ^{\sigma \dagger})_{\omega}   \nonumber \\
  &-& \sum_{\mu\nu}\sum_{\tau \sigma'} \frac{N_{\alpha\beta, \mu\nu}^{\sigma \sigma'}f_{\mu \tau}^{\sigma' *} \Gamma_{\sigma'}(-\omega - E_{\mu} + E_{\nu}) }{ \omega +E_{\alpha} - E_{\beta}}  G_{0}(A_{\tau \nu}|A_{\gamma \delta} ^{\sigma \dagger})_{\omega}   \nonumber \\
&-& \sum_{\mu\nu} \sum_{\lambda \tau \sigma'} \frac{N_{\alpha\beta, \mu\nu}^{\sigma \sigma'}L_{\mu \nu, \lambda \tau}^{\sigma'} \Lambda_{\sigma'}(-\omega - E_{\mu} + E_{\nu}) }{ \omega +E_{\alpha} - E_{\beta}}  G_{0}(A_{\lambda \tau}|A_{\gamma \delta} ^{\sigma \dagger})_{\omega}.   \nonumber \\
&&
\end{eqnarray}

At this stage, we will make a further approximation to $G_{1}(A_{\alpha \beta}|A_{\gamma \delta}^{\sigma \dagger})_{\omega}$ [Eq.(\ref{eq:92})] and $G_{2}(A_{\alpha \beta}|A_{\gamma \delta}^{\sigma \dagger})_{\omega}$ [Eq.(\ref{eq:94})]. In these equations, $G_{0}(A_{\alpha \beta}|A_{\gamma\delta}^{\sigma \dagger})_{\omega}$ appears on the level of $V_{k}^{2}$ and we can make simplifications which are exact at $V_k=0$,
\begin{eqnarray}  \label{eq:95}
  G_{0}(A_{\alpha \beta}|A_{\gamma \delta}^{\sigma \dagger})_{\omega} &=& \frac{\delta_{\alpha\gamma} \langle A_{\delta \beta} \rangle + \delta_{\beta \delta} \langle A_{\alpha \gamma} \rangle}{\omega + E_{\alpha} -E_{\beta}}  \nonumber \\
  & \approx & \delta_{\alpha\gamma}\delta_{\beta \delta} \frac{\langle A_{\alpha \alpha} \rangle + \langle A_{\beta \beta}  \rangle }{\omega + E_{\alpha} -E_{\beta}} .
\end{eqnarray}
That is, among the contributions higher than $V_{k}^{2}$, we neglect the processes that couple different SBOs and only consider the GFs that are diagonal on the BSO basis. With this approximation, Eqs.(\ref{eq:92}) and (\ref{eq:94}) are simplified greatly. Putting Eq.(\ref{eq:95}) into Eq.(\ref{eq:92}) and introduce the intermediate quantity similarly as in Eq.(\ref{eq:63}), one obtains
\begin{eqnarray}  \label{eq:96}
   \tilde{\Phi}_{\alpha \beta}^{\sigma}(\omega) &\equiv& \sum_{k} \frac{V_{k\sigma} \langle A_{\alpha \beta} c_{k \sigma} \rangle}{\omega - \epsilon_{k \sigma}}   \nonumber \\. 
 & \approx & \frac{f_{\beta \alpha}^{\sigma} \left( \langle A_{\alpha \alpha} \rangle + \langle A_{\beta \beta} \rangle \right)}{ \omega + E_{\beta} - E_{\alpha}} \varphi_{\beta \alpha}^{\sigma}(\omega).
\end{eqnarray}
$\varphi_{\beta \alpha}^{\sigma}(\omega)$ is given by Eq.(\ref{eq:64}). Comparing the obtained $G_{1}(A^{\sigma}_{\alpha \beta}|A_{\gamma \delta} ^{\sigma \dagger})_{\omega}$ with the bare EOM expansion result Eqs.(\ref{eq:67})-(\ref{eq:69}), we obtain
\begin{eqnarray}  \label{eq:97}
 &&  G_{1}(A^{\sigma}_{\alpha \beta}|A_{\gamma \delta} ^{\sigma \dagger})_{\omega} =  \frac{ \tilde{J}_{\alpha\beta, \gamma\delta}^{\sigma}(\omega) }{ \left( \omega + E_{\alpha} - E_{\beta} \right)\left( \omega + E_{\gamma} - E_{\delta} \right)}.  \nonumber \\
&& 
\end{eqnarray}

To simplify $G_{2}(A_{\alpha \beta}^{\sigma}|A_{\gamma \delta}^{\sigma \dagger})_{\omega}$, we put Eq.(\ref{eq:95}) into Eq.(\ref{eq:94}) and obtain
\begin{eqnarray}  \label{eq:98}
 &&  G_{2}(A^{\sigma}_{\alpha \beta}|A_{\gamma \delta} ^{\sigma \dagger})_{\omega} =  \frac{ \tilde{F}_{\alpha\beta, \gamma\delta}^{\sigma}(\omega) }{ \left( \omega + E_{\alpha} - E_{\beta} \right)\left( \omega + E_{\gamma} - E_{\delta} \right)}.  \nonumber \\
&& 
\end{eqnarray}
In the above two equations, $\tilde{J}_{\alpha\beta, \gamma\delta}^{\sigma}(\omega)$ and $\tilde{F}_{\alpha\beta, \gamma\delta}^{\sigma}(\omega)$ share the expression of $J_{\alpha\beta, \gamma\delta}^{\sigma}(\omega)$ (Eq.(\ref{eq:68})) and $F_{\alpha\beta, \gamma\delta}^{\sigma}(\omega)$ (Eq.(\ref{eq:69})), but with the substitution $a_{\alpha} \rightarrow \langle A_{\alpha \alpha} \rangle$ ($\alpha = 1 \sim 4$).
Note that the renormalized zeroth-order GF $G_{0}(A_{\alpha \beta}|A_{\gamma \delta}^{\sigma \dagger})_{\omega}$ in Eq.(\ref{eq:84}) is not changed and the averages there need to be calculated self-consistently.

Summing up Eqs.(\ref{eq:84}), (\ref{eq:97}), and (\ref{eq:98}), we obtain one of the simplest self-consistent schemes of EOM expansion for $G(A^{\sigma}_{\alpha \beta}|A_{\gamma \delta} ^{\sigma \dagger})_{\omega}$ which is exact to $V_{k}^{2}$,
\begin{eqnarray}  \label{eq:99}
 G(A^{\sigma}_{\alpha \beta}|A_{\gamma \delta} ^{\sigma \dagger})_{\omega}  &=&  \frac{\delta_{\alpha\gamma} \langle A_{\delta \beta} \rangle + \delta_{\beta \delta} \langle A_{\alpha \gamma} \rangle}{\omega + E_{\alpha} -E_{\beta}} \nonumber \\
 && + \frac{ \tilde{J}_{\alpha\beta, \gamma\delta}^{\sigma}(\omega) + \tilde{F}_{\alpha\beta, \gamma\delta}^{\sigma}(\omega) }{ \left( \omega + E_{\alpha} - E_{\beta} \right)\left( \omega + E_{\gamma} - E_{\delta} \right)}.  \nonumber \\
 &&
\end{eqnarray}
The averages of the type $\langle A_{\alpha \beta}\rangle$ in the above equation needs to be solved self-consistently from the full GF. Due to the same reason as discussed in the bare expansion, only averages of diagonal SBOs $\langle A_{\alpha \alpha}\rangle$ ($\alpha=1 \sim 4$) are involved in $G(d_{\sigma}|d_{\sigma}^{\dagger})_{\omega}$ . The advantage of the present self-consistent scheme is that it keeps the form of the bare GF expansion but only renormalizes the average values of the diagonal SBOs $A_{\alpha \alpha}$ ($\alpha=1 \sim 4$).

Let us now consider the self-consistent calculation of the remaining averages $\langle A_{\alpha \alpha} \rangle$ ($\alpha = 1 \sim 4$).
These averages need to be calculated from the CF-resummed GFs $\overline{G}_{CF}(A^{\uparrow}_{31}|A_{31} ^{\uparrow \dagger})_{\omega}$, $\overline{G}_{CF}(A^{\uparrow}_{24}|A_{24} ^{\uparrow \dagger})_{\omega}$, $\overline{G}_{CF}(A^{\downarrow}_{32}|A_{32} ^{\downarrow \dagger})_{\omega}$ and $\overline{G}_{CF}(A^{\downarrow}_{14}|A_{14} ^{\downarrow \dagger})_{\omega}$.
Under the particle-hole symmetry condition Eq.(\ref{eq:70}), Eq.(\ref{eq:99}) gives these GFs before resummation as
\begin{eqnarray}  \label{eq:100}
&& G(A^{\uparrow}_{31}|A_{31}^{\uparrow \dagger})_{\omega} = \frac{\tilde{W}_{1}^{\uparrow}}{\omega + U/2} + \frac{\tilde{W}_{3}^{\uparrow}(\omega)}{(\omega + U/2)^2};  \nonumber \\
&& G(A^{\uparrow}_{24}|A_{24}^{\uparrow \dagger})_{\omega} = \frac{\tilde{W}_{2}^{\uparrow}}{\omega - U/2} + \frac{\tilde{W}_{4}^{\uparrow}(\omega)}{(\omega - U/2)^2}; \nonumber \\
&& G(A^{\downarrow}_{32}|A_{32}^{\downarrow \dagger})_{\omega} = \frac{\tilde{W}_{1}^{\downarrow}}{\omega + U/2} + \frac{\tilde{W}_{3}^{\downarrow}(\omega)}{(\omega + U/2)^2}; \nonumber \\
&& G(A^{\downarrow}_{14}|A_{14}^{\downarrow \dagger})_{\omega} = \frac{\tilde{W}_{2}^{\downarrow}}{\omega - U/2} + \frac{\tilde{W}_{4}^{\downarrow}(\omega)}{(\omega - U/2)^2}. 
\end{eqnarray}
Their CF-resummations read
\begin{eqnarray}  \label{eq:101}
&& \overline{G}_{CF}(A^{\uparrow}_{31}|A_{31}^{\uparrow \dagger})_{\omega} = \frac{  (\tilde{W}_{1}^{\uparrow})^2 }{ \left(\omega + U/2\right) \tilde{W}_{1}^{\uparrow} - \tilde{W}_{3}^{\uparrow}(\omega) };  \nonumber \\
&& \overline{G}_{CF}(A^{\uparrow}_{24}|A_{24}^{\uparrow \dagger})_{\omega} = \frac{ (\tilde{W}_{2}^{\uparrow})^2 }{ \left(\omega - U/2 \right) \tilde{W}_{2}^{\uparrow} - \tilde{W}_{4}^{\uparrow}(\omega) }; \nonumber \\
&& \overline{G}_{CF}(A^{\downarrow}_{32}|A_{32}^{\downarrow \dagger})_{\omega} = \frac{ (\tilde{W}_{1}^{\downarrow})^2 }{ \left(\omega + U/2 \right)\tilde{W}_{1}^{\downarrow} - \tilde{W}_{3}^{\downarrow}(\omega) }; \nonumber \\
&& \overline{G}_{CF}(A^{\downarrow}_{14}|A_{14}^{\downarrow \dagger})_{\omega} = \frac{ (\tilde{W}_{2}^{\downarrow})^2 }{ \left(\omega - U/2 \right) \tilde{W}_{2}^{\downarrow} - \tilde{W}_{4}^{\downarrow}(\omega) }. 
\end{eqnarray}
The quantities $\tilde{W}_{i}^{\sigma}$ ($i=1 \sim 5$) in the above equations are given by 
\begin{eqnarray}  \label{eq:102}
 \tilde{W_{1}}^{\uparrow} &=& I_{13} ;    \nonumber \\
 \tilde{W_{2}}^{\uparrow} &=& I_{24} ;   \nonumber \\
  \tilde{W_{3}}^{\uparrow}(\omega)  &=& I_{14}\varphi_{14}^{\downarrow}(-\omega) - I_{23} \varphi_{32}^{\downarrow}(\omega)  \nonumber \\
  &&   + I_{13} \left[ \Gamma_{\uparrow}(\omega) +  \Lambda_{\downarrow}(\omega) - \Lambda_{\downarrow}(-\omega) \right];   \nonumber \\
  \tilde{W_{4}}^{\uparrow}(\omega) &=& I_{14} \varphi_{14}^{\downarrow}(\omega) - I_{23} \varphi_{32}^{\downarrow}(-\omega)   \nonumber \\
   && + I_{24} \left[ \Gamma_{\uparrow}(\omega) + \Gamma_{\downarrow}(\omega) - \Gamma_{\downarrow}(-\omega) - \Lambda_{\downarrow}(\omega) +  \Lambda_{\downarrow}(-\omega) \right];  \nonumber \\
  \tilde{W_{5}}^{\uparrow}(\omega) &=& - I_{14} \left[ \varphi_{14}^{\downarrow}(\omega) + \varphi_{14}^{\downarrow}(-\omega) \right] \nonumber \\
  && +  I_{23} \left[ \varphi_{32}^{\downarrow}(\omega) + \varphi_{32}^{\downarrow}(-\omega) \right]  \nonumber \\
 && + I_{24} \left[ -\Gamma_{\downarrow}(\omega) + \Gamma_{\downarrow}(-\omega) +\Lambda_{\downarrow}(\omega) - \Lambda_{\downarrow}(-\omega) \right]   \nonumber \\
 && + I_{13} \left[-\Lambda_{\downarrow}(\omega) + \Lambda_{\downarrow}(-\omega) \right].  
\end{eqnarray}
and
\begin{eqnarray}  \label{eq:103}
\tilde{ W_{1}}^{\downarrow} &=& I_{23} ;    \nonumber \\
\tilde{ W_{2}}^{\downarrow} &=& I_{14} ;   \nonumber \\
\tilde{ W_{3}}^{\downarrow}(\omega)  &=& I_{24}\varphi_{24}^{\uparrow}(-\omega) - I_{13} \varphi_{31}^{\uparrow}(\omega)  \nonumber \\
  &&   + I_{23} \left[ \Gamma_{\downarrow}(\omega) +  \Lambda_{\uparrow}(\omega) - \Lambda_{\uparrow}(-\omega) \right];   \nonumber \\
\tilde{W_{4}}^{\downarrow}(\omega) &=& I_{24} \varphi_{24}^{\uparrow}(\omega) - I_{13} \varphi_{31}^{\uparrow}(-\omega)   \nonumber \\
   && + I_{14} \left[ \Gamma_{\downarrow}(\omega) + \Gamma_{\uparrow}(\omega) - \Gamma_{\uparrow}(-\omega) - \Lambda_{\uparrow}(\omega) +  \Lambda_{\uparrow}(-\omega) \right];  \nonumber \\
\tilde{ W_{5}}^{\downarrow}(\omega) &=& I_{24} \left[ \varphi_{24}^{\uparrow}(\omega) + \varphi_{24}^{\uparrow}(-\omega) \right] \nonumber \\
  && -  I_{13} \left[ \varphi_{31}^{\uparrow}(\omega) + \varphi_{31}^{\uparrow}(-\omega) \right]  \nonumber \\
 && + I_{14} \left[ \Gamma_{\uparrow}(\omega) - \Gamma_{\uparrow}(-\omega) -\Lambda_{\uparrow}(\omega) + \Lambda_{\uparrow}(-\omega) \right]   \nonumber \\
 && + I_{23} \left[\Lambda_{\uparrow}(\omega) - \Lambda_{\uparrow}(-\omega) \right].  \nonumber \\
&&  
\end{eqnarray}
Here, $I_{\alpha \beta} = \langle A_{\alpha \alpha} \rangle + \langle A_{\beta \beta}\rangle$. The relation $\tilde{W}_{3}^{\sigma}(\omega) + \tilde{W}_{4}^{\sigma}(\omega) + \tilde{W}_{5}^{\sigma}(\omega) =\Gamma_{\sigma}(\omega)$ still holds. 
Compared to the same quantities $\{W_{i}^{\sigma}\}$ in the bare expansion Eq.(\ref{eq:72}) and Eq.(\ref{eq:73}) , the $1/2$ factor there is replaced with the averages $I_{\alpha\beta}$ in the renormalized expansion.
The self-consistent equations for the averages $\langle A_{\alpha \alpha} \rangle$ ($\alpha = 1 \sim 4$) are completed by 
\begin{equation}  \label{eq:104}
   \langle A_{\beta \beta} \rangle = -\frac{1}{\pi} \displaystyle\int_{-\infty}^{\infty} {\text Im} \overline{G}_{CF}(A^{\sigma}_{\alpha \beta}|A_{\alpha \beta}^{\sigma \dagger})_{\omega + i \eta} \frac{1}{e^{\beta \omega} + 1 } d\omega. 
\end{equation}
and $\sum_{\beta} \langle A_{\beta \beta} \rangle = 1$.

With the averages $\langle A_{\alpha \alpha} \rangle$ ($\alpha=1 \sim 4$) obtained, the single particle GFs $G(d_{\uparrow}|d_{\uparrow}^{\dagger})_{\omega} = G(A_{31}^{\uparrow} + A_{24}^{\uparrow}|A_{31}^{\uparrow \dagger} + A_{24}^{\uparrow \dagger} )_{\omega}$ and $G(d_{\downarrow}|d_{\downarrow}^{\dagger})_{\omega} = G(A_{32}^{\downarrow} - A_{14}^{\downarrow}|A_{32}^{\downarrow \dagger} - A_{14}^{\downarrow \dagger} )_{\omega}$ are calculated as
\begin{eqnarray}  \label{eq:105}
   G(d_{\sigma}|d_{\sigma}^{\dagger})_{\omega} &=&   \frac{\tilde{W}_{1}^{\sigma}}{\omega + U/2} + \frac{\tilde{W}_{2}^{\sigma}}{\omega - U/2}  + \frac{\tilde{W}_{3}^{\sigma}(\omega)}{ (\omega + U/2)^2 }  \nonumber \\
  && +\frac{\tilde{W}_{4}^{\sigma}(\omega)}{ (\omega - U/2)^2 }  + \frac{\tilde{W}_{5}^{\sigma}(\omega)}{(\omega + U/2)(\omega - U/2) }. \nonumber \\
 &&   
\end{eqnarray}
The CF resummation is then carried out to it in the same way as for the bare GF expansion Eq.(\ref{eq:80}) and (\ref{eq:81}). The result is denoted as $\overline{G}_{SC}(d_{\sigma}|d_{\sigma}^{\dagger})_{\omega}$ and we obtain
\begin{equation}   \label{eq:106}
 \overline{G}_{SC}(d_{\sigma}|d_{\sigma}^{\dagger})_{\omega}  
 = \cfrac{a_0}{\omega + b_1
          - \cfrac{a_1}{\omega + b_2 } }, 
\end{equation}
with coefficients
\begin{eqnarray}    \label{eq:107}
&&  a_{0} = 1;  \nonumber \\
&&  a_{1} = \left(U/2 \right)^2;  \nonumber \\
&&  b_{1} = \frac{U}{2} (\tilde{W}_{1}^{\sigma} - \tilde{W}_{2}^{\sigma}) - \Gamma_{\sigma}(\omega);   \nonumber \\
&&  b_{2} = -\frac{U}{2}(\tilde{W}_{1}^{\sigma} - \tilde{W}_{2}^{\sigma}) - \Gamma_{\sigma}(\omega) + 2\tilde{W}_{5}^{\sigma}(\omega).
\end{eqnarray}
To obtain these results, we have made use of the fact that $\tilde{W}_{3}^{\sigma} - \tilde{W}_{4}^{\sigma} \propto V_{k}^4$ and neglected them when comparing the $1/\omega$ expansion of $G(d_{\sigma}|d_{\sigma}^{\dagger})_{\omega}$ and $\overline{G}_{SC}(d_{\sigma}|d_{\sigma}^{\dagger})_{\omega}$. For the paramagnetic bath, Eqs.(\ref{eq:106}) and (\ref{eq:107}) reduce to  Eqs.(\ref{eq:80}) and (\ref{eq:81}) of the bare EOM expansion with CF-resummation.
Note that the inverse order: First doing CF-resummation for $G(A_{\alpha \beta}|A_{\gamma \delta}^{\sigma \dagger})_{\omega}$ and then summing them up to produce $G(d_{\sigma}|d_{\sigma}^{\dagger})_{\omega}$, does not work. This is because some components, e.g., $G(A_{31}|A_{24}^{\dagger})_{\omega}$ and $G(A_{24}|A_{31}^{\dagger})_{\omega}$, starts from $1/\omega^2$ in the $1/\omega$ expansion and the form of CF Eq.(\ref{eq:106}) does not apply.

\subsection{Numerical Results}
In this subsection, we present numerical results for the formula obtained in previous subsections. We compare results obtained from the three different combinations of second-order strong-coupling expansions and resummation methods: bare EOM expansion with SE resummation (bSE), bare EOM expansion with CF resummation (bCF), and self-consistent EOM expansion with CF resummation (SC). All these results are compared with the numerical renormalization group (NRG) data, which is believed to be accurate at the low and small-frequency regimes.
\begin{figure}[t!]   
\begin{center}
\includegraphics[width=5.5in, height=4.4in, angle=0]{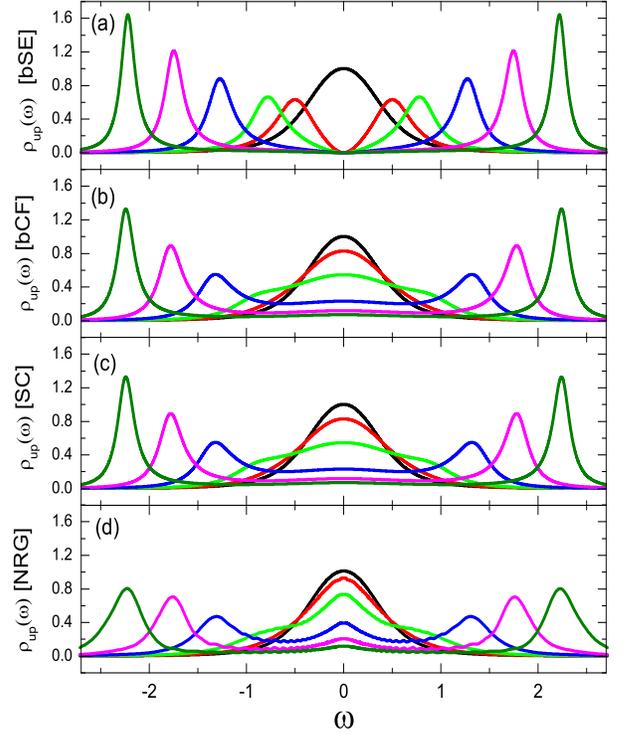}
\vspace*{-0.5cm}
\end{center}
\caption{(Color online) The impurity density of states $\rho_{\uparrow}(\omega$ for various $U$'s, obtained using (a) bSE; (b) bCF; (c) SC; and (d) NRG. From top to bottom at small the $\omega$ regime: $U=0.0, 0.5, 1.0, 2.0, 3.0, 4.0$, respectively. Other model parameters are $\Gamma=0.1$, $\Delta\omega=0.0$, $\epsilon_d=-U/2$, and $T=0.1$. NRG parameters are $\Lambda=3.0$, $M_s = [256, 280]$, log-Gaussian broadening parameter $B=0.08$, and $N_z=8$.} 
\end{figure}
\begin{figure}[t!]    
\vspace{-2.0cm}
\begin{center}
\includegraphics[width=5.9in, height=4.3in, angle=0]{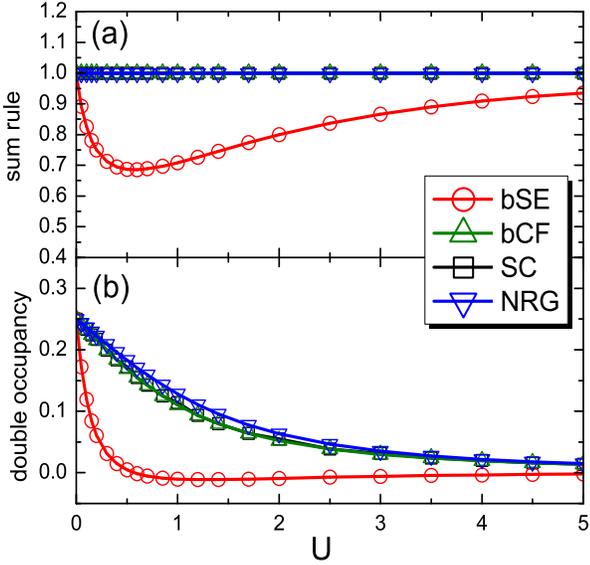}
\vspace*{-1.0cm}
\end{center}
\caption{(Color online) (a) The sum rule of impurity density of states as functions of $U$ in paramagnetic bath; (b) double occupancy as functions of $U$. The symbols with guiding lines are bSE (circles), bCF (up triangles), SC (squares), and NRG (down triangles). The model parameters are $\Gamma=0.1$, $\Delta\omega=0.0$, $\epsilon_d=-U/2$, and $T=0.1$. NRG parameters are same as in Fig.1 but without the $z$-average.} 
\end{figure}
We use a Lorentzian hybridization function for the Anderson impurity model,
\begin{equation}    \label{eq:108}
    \Delta_{\sigma}(\omega) = \frac{\Gamma \omega_{c}^2}{ (\omega + \sigma \Delta \omega)^2 + \omega_{c}^2 } . 
\end{equation}
Here, $\omega_c = 1.0$ is the energy unit. $\Gamma$ is the hybridization strength. The spin-dependent energy shift $\Delta \omega$ is introduced to simulate the spin-dependent bath energies. $\sigma = +1$ for spin up and $\sigma = -1$ for spin down. $\Delta \omega =0$ gives paramagnetic bath while $\Delta \omega \neq 0$ mimics the spin-polarized bath. In this paper, we only study the particle-hole symmetric point $\epsilon_d = -U/2$.

NRG calculation of the local density of states (LDOS) for this Anderson impurity model is done with the full density matrix formalism,\cite{Weichselbaum1} supplemented with the SE trick.\cite{Bulla1} To discern the sharp Hubbard band in the large-$U$ regime, we use a small broadening parameter and average the LDOSs with $N_z=8$ interleaved discretizations.\cite{Yoshida1} The final LDOS from this standard procedure is believed to be accurate at least for low temperature and in small frequencies. The thermodynamical quantities $\langle n_{\sigma}\rangle$ and $\langle n_{\uparrow} n_{\downarrow}\rangle$ are obtained from the respective spectral function by frequency integration.

Figure 1 presents LDOS at $T=0.1$ and various $U$'s for the paramagnetic bath $\Delta \omega =0$. The data are for bSE, bCF, SC, and NRG in Figs.1(a), 1(b), 1(c), and 1(d), respectively. Among the three combinations of expansion-resummation method, LDOS of bSE has an unphysical dip at $\omega=0$ for any non-zero $U$. bCF and SC give out identical LDOS because SC reduces to bCF in the particle-hole symmetric and paramagnetic situation. The height of the central peak obtained from bCF [Fig.1(b)] and SC [Fig.1(c)] decreases gradually with increasing $U$. This behavior, being consistent with NRG, is correct for temperature higher than the Kondo temperature. In all the obtained data, the Hubbard peak positions are slightly larger than $U/2$ due to the hybridization shift. In the large $U$ limit, all LDOSs tend to the atomic form $\rho_{at}(\omega) = 1/2 \delta(\omega - U/2) + 1/2 \delta(\omega + U/2)$ which is expected when $U$ is much larger than bath band width. Compared to the NRG curve in Fig.1(d), qualitative agreement is reached by bCF and SC in both the small- and large-$U$ limits. The deviation is stronger in the small-$\omega$ regime for intermediate $U$ values $1.0 \leq U \leq 3.0$.

The good quality of LDOS at small $U$ obtained from strong-coupling expansion is a consequence of resummation which effectively extends the validity range of the series expansion. Actually, all the three expansion-resummation schemes give exact GF at $U=0$.
 It is observed that for large $U$ values, the Hubbard peaks are significantly sharper than NRG results. The neglecting of higher-order contributions of hybridization may lead to sharper Hubbard peaks, but we believe that the main reason for this discrepancy is the poor energy resolution of NRG at high energies. It is known that NRG tends to over broaden high energy peaks. Indeed, the height of Hubbard peaks increases when we use smaller broadening parameter, larger $N_z$, and keep more states. Here we have used the log-Gaussian broadening. Sharp features have been obtained from Gaussian instead of log-Gaussian broadening and $N_{Z}$ up to $32$.\cite{Zitko1}

Figure 2 presents the sum rule and double occupancy as functions of $U$ at $T=0.1$ and $\Delta \omega = 0.0$. In Fig.2(a), integration of the LDOS in Fig.1 are compared. As expected, LDOS from bSE does not obey the sum rule, while those from bCF and SC obey it at a precision $10^{-4}$. The tiny deviation is due to numerical error. NRG result fulfills the sum rule at machine precision.

The double occupancy $\langle n_{\uparrow}n_{\downarrow} \rangle$ can be calculated in various ways. For bSE and bCF, it can be calculated directly from the single particle GF as
\begin{eqnarray}    \label{eq:109}
 &&  \langle n_{\uparrow}n_{\downarrow} \rangle = -\frac{1}{\pi} \displaystyle\int_{-\infty}^{\infty} {\text Im} G(n_{\bar{\sigma}}d_{\sigma}| d_{\sigma}^{\dagger})_{\omega + i\eta} \frac{1}{e^{\beta \omega} + 1} d\omega;   \nonumber \\
 &&  G(n_{\bar{\sigma}}d_{\sigma}| d_{\sigma}^{\dagger})_{\omega} = \frac{1}{U} G(d_{\sigma}| d_{\sigma}^{\dagger})_{\omega} \Sigma_{\sigma}(\omega).  
\end{eqnarray}
One can use either $\sigma=\uparrow$ or $\sigma=\downarrow$ in the above equations. For SC, besides the above equation, one could also use $\langle n_{\uparrow}n_{\downarrow} \rangle = \langle A_{44} \rangle$ since $\langle A_{44} \rangle$ has been obtained in the self-consistent calculation. Different ways of calculating $\langle n_{\uparrow}n_{\downarrow} \rangle$ have relative deviations smaller than $5\%$. In this paper, for bSE, bCF and NRG, we use Eq.(\ref{eq:109}) with $\sigma=\uparrow$ and for SC we use $\langle n_{\uparrow}n_{\downarrow} \rangle = \langle A_{44} \rangle$.

In Fig.2(b), the bSE result for the double occupancy is much smaller than the other three, and even slightly negative near $U=1.0$. The results of bCF and SC agree with that of NRG at quantitative level. Similarly to LDOS, the agreement of the bCF and SC results with NRG is better in both the small- and large-$U$ regimes.
\begin{figure}[t!]    
\begin{center}
\includegraphics[width=5.5in, height=4.4in, angle=0]{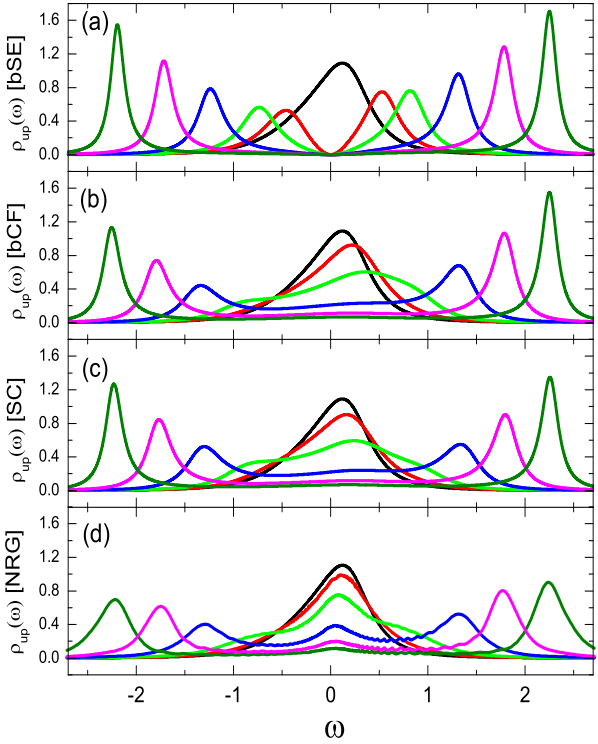}
\vspace*{-1.0cm}
\end{center}
\caption{(Color online) The impurity density of states $\rho_{\uparrow}(\omega$ for various $U$'s in a spin polarized bath, obtained using (a) bSE; (b) bCF; (c) SC; and (d) NRG. From top to bottom at the small-$\omega$ regime: $U=0.0, 0.5, 1.0, 2.0, 3.0, 4.0$, respectively. Other model parameters are $\Gamma=0.1$, $\Delta\omega=0.2$, $\epsilon_d=-U/2$, and $T=0.1$. NRG parameters are same as in Fig.1.} 
\end{figure}
\begin{figure}[t!]    
\vspace{-1.5cm}
\begin{center}
\includegraphics[width=5.8in, height=4.6in, angle=0]{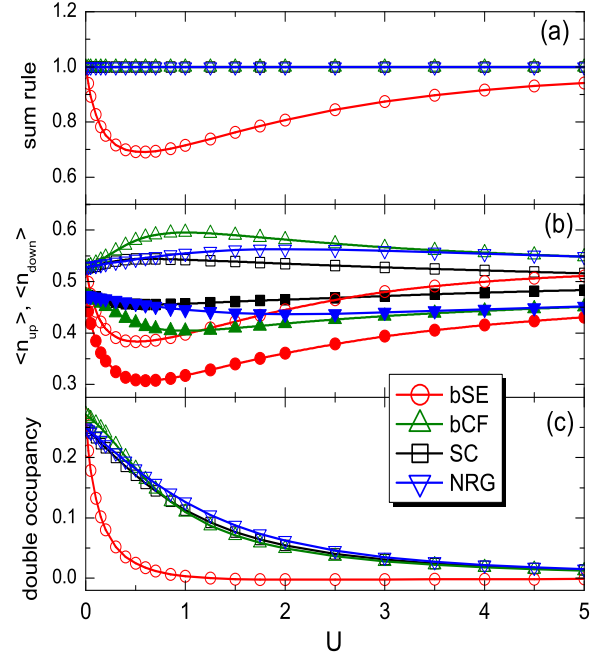}
\vspace*{-1.5cm}
\end{center}
\caption{(Color online) Physical quantities as functions of $U$ at fixed $T=0.1$ for the Anderson impurity model with spin polarized bath. (a) Sum rule; (b) averages $\langle n_{\uparrow} \rangle$ (filled symbols) and $ \langle n_{\downarrow} \rangle$ (empty symbols); and (c) double occupancy. The symbols with guiding lines are bSE (circles), bCF (up triangles), SC (squares), and NRG (down triangles). Model parameters are $\Gamma=0.1$, $\Delta\omega=0.2$, $\epsilon_d=-U/2$. NRG parameters are same as Fig.1 but without the $z$-average.} 
\end{figure}

In Fig.3 and Fig.4, we focus on the Anderson impurity model with a spin-polarized bath $\Delta\omega = 0.2$, all at an intermediate temperature $T=0.1=\Gamma$. In Fig.3, the $U$-dependence of LDOS is shown. Since $\rho_{\downarrow}(\omega) = \rho_{\uparrow}(-\omega)$ is obeyed very well, here we only show $\rho_{\uparrow}(\omega)$. All the LDOS curves have an asymmetric shape due to spin polarization in the bath. Similar to Fig.1(a), bSE curves have unphysical dips at $\omega = 0$ for nonzero $U$. the bCF and SC results are similar, but no longer identical in the case of magnetic bath. It is seen that the bCF result has more asymmetry in the upper and lower Hubbard bands than does the SC result.

In Fig.4(a), the sum rule of LDOSs is analyzed. bSE has an incorrect sum rule while bCF and SC fulfill it perfectly. In Fig.4(b), the impurity electron occupancies $\langle n_{\uparrow} \rangle$ (filled symbols) and $\langle n_{\downarrow} \rangle$ (empty symbols) are shown as functions of $U$. Due to the incorrect sum rule, bSE gives a total occupancy less than half-filling. In contrast, $\langle n_{\uparrow} \rangle -1/2 = - ( \langle n_{\downarrow}\rangle -1/2)$ is preserved in the results of bCF, SC and NRG. Being consistent with the larger asymmetry in LDOS, bCF gives qualitatively larger magnetization $M = |\langle n_{\uparrow} \rangle - \langle n_{\downarrow}\rangle |$ than SC. 
Compared to NRG data, the bCF result agrees better for $U \geq 3.0$ while the SC result has better behavior for $U \leq 1.0$ and all curves are non-monotonic. The SC result for $M$ is appreciably smaller than NRG even in $U=5.0$. 

This quantitative difference can be traced back to the approximation scheme Eq.(\ref{eq:93}) and (\ref{eq:95}) used for the self-consistent calculation of averages. In those equations, the atomic-like truncation scheme weakens the influence of the asymmetric bath on the impurity and leads to smaller $M$. In other words, although the single particle GF being exact up to $V_{k}^{2}$, the averages are actually evaluated with respect to a ground state or a density matrix which is accurate to lower order. As will be detailed later, the significant error in the large-$U$ regime hints that not only $\Gamma/U$, but also $\Gamma/T$ should be small to guarantee quantitative accuracy in the present expansion scheme.

In Fig.4(c), the double occupancies are shown as functions of $U$. They look similar to the paramagnetic case while the bCF result has some drawback in the magnetic case: $\langle n_{\uparrow} n_{\downarrow} \rangle$ exceeds the upper limit $1/4$ in the small $U$ limit. This reflects that although the single particle GF obtained from bCF is exact at $U=0$, the two-particle GF is not. When $U$ is small, something is qualitatively wrong in the higher order GFs obtained from bCF. In contrast, SC result agrees with NRG much better from small to large $U$.

\begin{figure}[t!]  
\vspace*{-1.0cm}
\begin{center}
\includegraphics[width=5.5in, height=4.4in, angle=0]{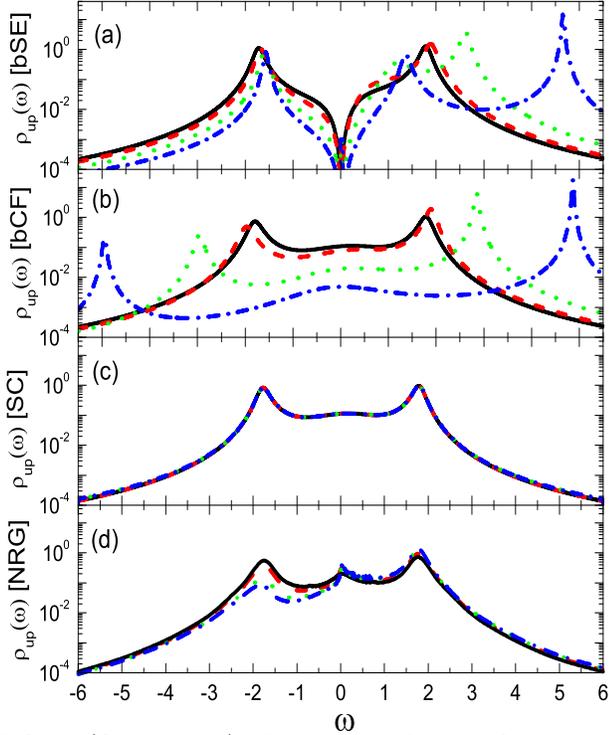}
\vspace*{-1.5cm}
\end{center}
\caption{(Color online) The impurity density of states $\rho_{\uparrow}(\omega)$ for $U=3.0$ and various $T$'s in a spin polarized bath, obtained using (a) bSE; (b) bCF; (c) SC; and (d) NRG. The curves are for $T=0.1$ (solid line), $T=0.03$ (dashed line), $T=0.01$ (dotted line), and $T=0.005$ (dash-dot line). Other model parameters are $\Gamma=0.1$, $\Delta\omega=0.2$, and $\epsilon_d=-U/2$. NRG parameters are same as in Fig.1.}  
\end{figure}
\begin{figure}[t!]    
\vspace{-1.5cm}
\begin{center}
\includegraphics[width=5.8in, height=4.6in, angle=0]{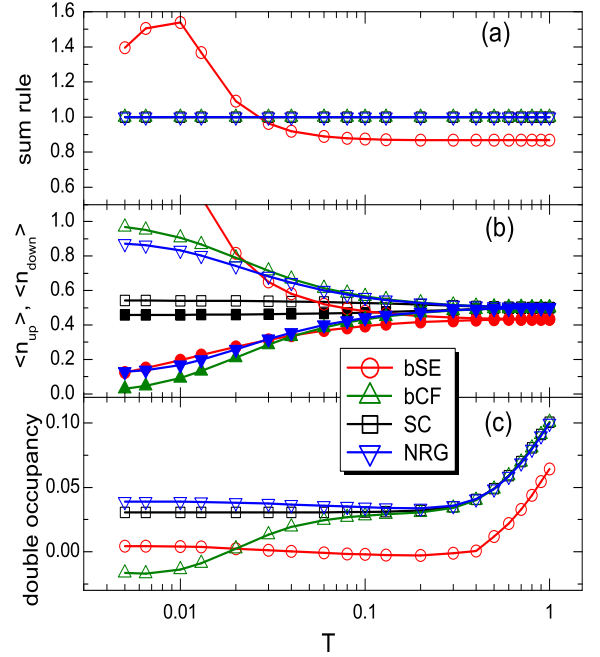}
\vspace*{-1.5cm}
\end{center}
\caption{(Color online) Physical quantities as functions of $T$ in log scale at $U=3.0$ for the Anderson impurity model with spin polarized bath. (a) Sum rule; (b) averages $\langle n_{\uparrow} \rangle$ (filled symbols) and $ \langle n_{\downarrow} \rangle$ (empty symbols); and (c) double occupancy. The symbols with guiding lines are bSE (circles), bCF (up triangles), SC (squares), and NRG (down triangles). Model parameters are $\Gamma=0.1$, $\Delta\omega=0.2$, $\epsilon_d=-U/2$. NRG parameters are same as in Fig.1 but without the $z$-average.} 
\end{figure}

To study the temperature dependence of these results, in Fig.5 we show LDOS on log-scale at $U=3.0$, $\Delta \omega = 0.2$ for various temperatures. In Fig.5(a), it is seen that LDOS from bSE are normal at $T=0.1$ (black solid line) (the dip at $\omega=0$ still present). As temperature lowers further, the lower Hubbard band stays around $-U/2$ while the upper Hubbard band begins to split and the higher branch moves to larger values. Figure 5(b) shows the LDOS from bCF. In the limit $T=0$, both the lower and the upper Hubbard bands move towards $\pm \infty$, with enhanced weight transfer from the lower Hubbard band to the higher one. Numerically we find that the position of these peaks is proportional to $1/T$ in the small $T$ limit. They are the unphysical features due to the $\beta$ factors in the bare expansion of GF. After either bSE or bCF resummation, these factors enter the denominator and influence the position of poles. In the second-order bare EOM expansion Eq.(\ref{eq:71}), it is $W_{1}^{\sigma}$ and $W_{2}^{\sigma}$ that contain the $\beta$ factor via $\langle A_{\alpha\alpha}\rangle_{2}$  ($\alpha = 1 \sim 4$). For the paramagnetic bath, $W_{1}^{\sigma}=W_{2}^{\sigma}=1/2$, the $\beta$-dependent terms cancel and the problem does not appear. For the magnetic bath, we numerically find that $\langle A_{11} \rangle_{2}$ and $\langle A_{22} \rangle_{2}$ are proportional to $\beta$, while $\langle A_{33} \rangle_{2}$ and $\langle A_{44} \rangle_{2}$ are almost independent of $\beta$. 

The LDOS from SC shown in Fig.5(c) has very weak temperature dependence. As $T$ decreases from $T=0.1$, the Hubbard peak positions move weakly and converge to $\pm 1.8$. The weight distribution does not change much down to $T=0.005$. Compared to the NRG results in Fig.5(d), SC gives quantitatively correct peak position of Hubbard bands. What is missing in the SC results is the central Kondo resonance at small $T$ and the continuous weight transfer from the lower Hubbard band to the higher one as $T$ decreases. This is again a consequence of the atomic-like truncation scheme used in Eq.(\ref{eq:93}) and Eq.(\ref{eq:95}). Overall, in the low temperature limit, bSE and bCF have diverging positions of Hubbard peaks, while SC maintains correct peak position but has weaker spectral weight transfer compared to NRG. In the high temperature regime $T \gg \Gamma$, the shape of LDOS from bCF and SC agree well with NRG.

In Fig.6, the sum rule, electron occupation, and the double occupancy as functions of $T$ are presented. The sum rule in Fig.6(a) shows that bSE result is incorrect for all temperatures, while bCF and SC results keep at unity, as expected. The electron occupations in Fig.5(b) show significant difference among the four results at low $T$. The result of bSE does not fulfill particle-hole symmetry and $\langle n_{\downarrow} \rangle$ exceeds unity. bCF and SC give qualitatively correct results. At the low temperature limit, bCF gives out a fully polarized impurity $M=1.0$, consistent with the enhanced asymmetry in LDOS at the low-$T$ regime. In contrast, SC gives much weaker polarization, with $M$ saturating to $0.084$ at $T=0.005$, much smaller than the NRG value $0.742$. At the high-temperature regime $T \gg \Gamma$, bCF and SC results agree well with NRG.

Double-occupancy results are shown in Fig.6(c). The result from bSE is much smaller than the others and slightly negative around $T=0.4$. At low temperatures, bCF produces a qualitatively wrong result: $\langle n_{\uparrow} n_{\downarrow}\rangle$ tends to negative for $T \leq 0.02$. In contrast, the SC result decreases with lower $T$ at high temperatures and reaches a constant $\langle n_{\uparrow} n_{\downarrow}\rangle = 0.03$ for $T \leq 0.4$, close to the NRG value $0.039$ at the low-temperature limit. At high temperatures, the double occupancy from bCF and SC agree well with that of NRG.

\section{Discussion and Summary}

\begin{figure}[t!]    
\vspace{-0.8in}
\begin{center}
\includegraphics[width=6.4in, height=5.3in, angle=0]{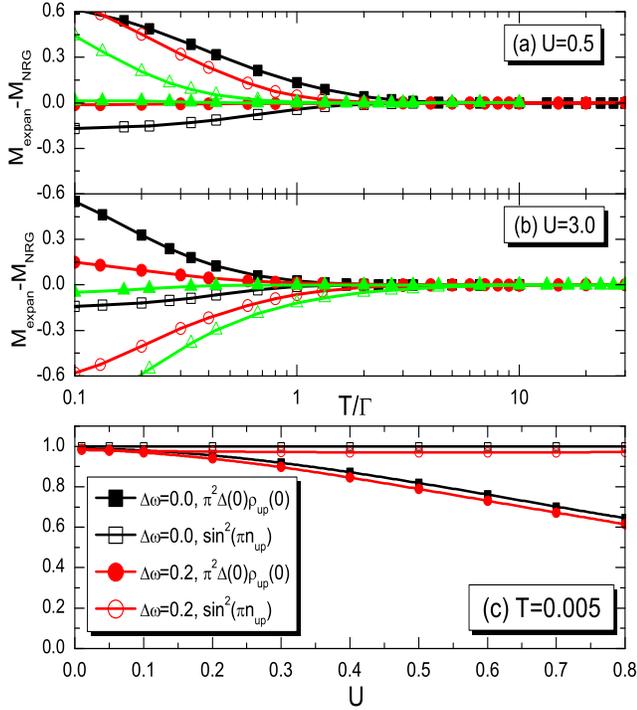}
\vspace{-1.5cm}
\end{center}
\caption{(Color online) (a) and (b): The differences between the impurity magnetization $M= |\langle n_{\uparrow}- n_{\downarrow} \rangle |$ obtained from expansions and NRG, as functions of $T$. The parameters are (a) $U=0.05$ and (b) $U=3.0$, at $\Delta\omega=0.2$. The filled symbols are for bCF and the empty symbols are for SC. Squares, circles, and triangles correspond to $\Gamma=0.03$, $\Gamma=0.1$, and $\Gamma=0.3$, respectively. (c) Checking Friedel sum rule at $T=0.005$ and $\Gamma=0.1$ for small $U$ values. Filled symbols are $\pi^{2}\Delta_{\uparrow}(0) \rho_{\uparrow}(0)$ and empty symbols are $\text{sin}^{2}(\pi n_{\uparrow})$.} 
\end{figure}

We first discuss the validity range of the expansion schemes. The quality of an expansion is controlled by the small parameter $\lambda$ in $H$, which in our strong-coupling expansion for the  Anderson impurity model is characterized by the hybridization strength $\Gamma$. Roughly speaking, the validity range of the expansion is $\Gamma$ much smaller than all the other energy scales, including the interaction $U$, temperature $T$, and frequency $\omega$. In Figs.1 $\sim$ 4, $T = \Gamma=0.1$ is used which is on the boundary of the validity range, and the results have significant errors. It is seen from Fig.6 that the agreement between the expansion and NRG is good for $T \gg \Gamma$ and $U \gg \Gamma$. Figures 3 and 4 shows that the agreement is also good for small $U \sim 0$ due to the $U^{1}$-order accuracy acquired from the CF-resummation. 

Figure 7(a) and 7(b) demonstrate that $\Gamma$ indeed sets in as the breakdown temperature scale in the present strong-coupling expansion approaches. Taking the impurity magnetization $M=\langle n_{\uparrow} \rangle - \langle n_{\downarrow} \rangle$ as an example, we show the difference between the expansion and NRG results $M_{exp} - M_{NRG}$ as functions of $T/\Gamma$, for $U=0.5 \sim \Gamma$ in Fig.7(a) and for $U=3.0 \gg \Gamma$ in Fig.7(b). For $\Gamma$ values ranging from $0.03$ to $0.3$, the magnetization from both bCF and SC approach the NRG values at $T/\Gamma \gg 1$, and deviate significantly at $T/\Gamma \leq 1$. Similar behavior is observed in the double occupancy curve (not shown here). In Fig.7(c), we examine to what extent the Friedel sum rule $\pi^{2}\Delta_{\sigma}(0)\rho_{\sigma}(0) = \text{sin}^{2}(\pi n_{\sigma})$ is satisfied at low temperatures.\cite{Weichselbaum1} It is seen that for both spin polarized and un-polarized cases, this relation is fulfilled exactly only at $U=0$. Approximate fulfillment in seen in $U < \Gamma = 0.1$.  Since the Friedel sum rule is a consequence of the Fermi liquid ground state, it is not expect to be fulfilled well by the strong-coupling expansion which starts from the local moment limit. However, due to the CF-resummation method used in our approach,  GF acquires the correct $U$ term and the Friedel sum rule is satisfied in the regime $U < \Gamma$.

To summarize, for $T \gg \Gamma$, both bCF and SC schemes are quantitatively accurate in the regime $U \gg \Gamma$, and produce a smoothly interpolation between large $U$ and small $U$. For $T \leq \Gamma$, accurate results can only be expected for $U \sim 0$. Since the Kondo scale $T_{K}$ is much smaller than $\Gamma$ for large $U$, the Kondo resonance cannot be described by the the present strong-coupling expansion method. 

Although the SC scheme improves over bSE and bCF on the causality and the zero-temperature divergence problems, the appearance of a breakdown scale $\Gamma$ in the temperature axis is an obvious shortcoming, similar to the strong-coupling expansion for the Hubbard model.\cite{Pairault1,Pairault2} Considering that for the Anderson impurity model, much better results are available from methods such as NRG, QMC, and functional renormalization group method\cite{Karrasch1}, the present expansion schemes receives only partial success at this stage. However, the results are amenable to further improvement. The appearance of the breakdown scale in temperature and the zero-temperature divergence problem are related to the fact that $V_{k}=0$ is a singular point in the ground state of $H$ with a spin-polarized bath. One could avoid these problems by selecting a suitable $H_0$ whose ground state is continuously connected to that of $H$. This issue will be studied in the future.

Below, we discuss some distinct features of the present expansion method compared to existing theories. The present approach is universal in the sense that it has no requirement on the form of $H_0$. For any $H_0$ that is exactly solvable, i.e., either its GF EOM closes naturally, or its eigen states and eigen values are obtainable, series expansion of GF can be constructed in a unified framework. $H_0$- and $H_1$-specific diagrammatic rules are not needed in this method. By using the self-consistent EOM expansion supplemented with CF resummation, causality of GF is guaranteed and the zero-temperature divergence problem removed. The resulting GF has extended range of validity. Therefore, for those Hamiltonians that both $H_0$ and $H_1$ are exactly solvable, by expanding GF from the two limits and comparing the results, one can obtain reliable knowledge in both the small and large interaction regimes. In principle, expansion around a cluster or impurity Hamiltonian $H_0$ is also possible. This could provide a possible alternative derivation of the cluster perturbation theory\cite{Senechal1} or dual fermion dynamical mean-field theory.\cite{Rubtsov1}

The present approach is distinctive in that arbitrary double time GF can be expanded in the same framework. In the traditional methods, calculation of GFs of more than one particle is a laborious task. Here, due to the universality of the formalism, multiple-particle GF can be expanded in a way parallel to one-particle GF. For an example, the strong-coupling expansion method used in this paper for calculating the single particle GF can also be used to produce the dynamical spin- or charge- correlation functions, with only slight modification in the formalism. 

Next, we discuss possible improvement and extensions of the present approach. As seen in Fig.4(b) and Fig.6(b), $\langle n_{\sigma}\rangle$ from the second-order self-consistent strong-coupling expansion deviates significantly from the NRG result, even for $U$ as large as $5.0$. Also, the temperature dependence of $M$ is too weak. Obviously there is much room to improve the result. To calculate the averages of the type $\langle A_{\gamma\mu}^{\sigma \dagger} c_{k \sigma'} \rangle$ (in Eq.(\ref{eq:88})) and $\langle A_{\lambda \gamma} c_{p \sigma''} c_{k \sigma'} \rangle$ [in Eqs.(\ref{eq:90}) and (\ref{eq:91})], instead of using the atomic-like truncation scheme, we can carry out the self-consistent EOM expansion for GFs $G(c_{k \sigma'}|A_{\gamma\mu}^{\sigma \dagger})_{\omega}$ and $G( c_{k \sigma'}| A_{\lambda \gamma} c_{p \sigma''})_{\omega}$ to $V_{k}^{2}$ order and calculate the averages from the CF-resummed GFs. Also, a suitable selection of $H_0$ may help remedy the zero-temperature divergence problem and remove the breakdown scale in temperature.

A by-product of the present method is the EOM for the $n$-th order residue $\Gamma_{n}(A|B)_{\omega}$, Eq.(\ref{eq:11}) or Eq.(\ref{eq:25}). It could be employed to produce higher order modifications to the series up to $G_{n}(A|B)_{\omega}$. The EOM of $\Gamma_{n}(A|B)_{\omega}$ is formally similar to that of the full GF $G(A|B)_{\omega}$, except that the lower order contributions have been singled out. In principle, it can be solved approximately by standard truncation schemes. This provide possibilities of constructing new types of truncation approximations which are exact up to order $\lambda^n$, or developing improved CF resummation formulas with a terminator.\cite{Hayn1}

An ideal expansion scheme may be that the resulting GF is exact simultaneously to $V_{k\sigma}^2 U^{\infty}$ and $V_{k\sigma}^{\infty} U^{2}$, and hence accurate in both the weak- and strong-coupling limits. As it is difficult to realize in traditional perturbation theories, it is apparently possible to achieve this goal in the EOM expansion method. One could first carry out the weak-coupling expansion to obtain $G^{(2)}(d_{\sigma}|d_{\sigma}^{\dagger})_{\omega}$, and then carry out the strong-coupling expansion to $V_{k}^2$ order for the residue $\Gamma_{2}(d_{\sigma}|d_{\sigma}^{\dagger})_{\omega}$ by employing its EOM. The resulting GF $G^{(2,2)}(d_{\sigma}|d_{\sigma}^{\dagger})_{\omega}$ satisfies the above requirement. In practice, however, the resummation method suitable for such an expansion is yet to be developed.

Direct multiple-variable expansion is also possible within the framework of EOM expansion. For an example, the splitting of Hamiltonian $H= H_0 + \lambda H_1 + \theta H_2$ can be used to generate a GF expansion $G(A|B)_{\omega} = \sum_{i,j}\lambda^{i}\theta^{j} G_{ij}(A|B)_{\omega}$. If we choose $H_0 = \sum_{k\sigma} \epsilon_{k \sigma} c_{k \sigma}^{\dagger} c_{k\sigma}$, $H_1 = \sum_{k \sigma} V_{k\sigma} ( c_{k\sigma}^{\dagger} d_{\sigma} + d_{\sigma}^{\dagger}c_{k\sigma} )$, and $H_2 = U n_{\uparrow} n_{\downarrow} +\epsilon_d \sum_{\sigma} n_{\sigma}$, the expansion up to lowest several orders can be obtained with ease. A subsequent resummation can be used to produce a physically meaningful result, being correct in both the strong- and weak-coupling limits. If the full Hamiltonian is treated as a perturbation, the self-consistent expansion will produce a moment expansion of GF, while the bare expansion is equivalent to the simultaneous moment and high temperature expansions.

The same strategy of expanding the double time GF can be extended straightforwardly to other GFs, if only the EOM formalism also applies there. For an example, the Keldysh GF describing the non-equilibrium process can be described by EOM. The present EOM-based expansion method can be extended to calculate the the Keldysh GF.

From our demonstrative calculation for the weak- as well as strong-coupling expansion, it is clear that the present method also has some shortcomings. For most of the models, it is difficult to obtain explicit expansion higher than second order because the complexity of calculation increases very fast with order. This feature is common in every expansion method such as the Feynman diagram for weak-coupling expansion, Metzner's diagram for strong-coupling expansion,\cite{Metzner1} or Dai's direct expansion.\cite{Dai1} In these techniques, the time ordering and multiple integrals will complicate the problem. With the aid of computer algebra, we hope that higher order GF could be obtained, similarly to the situation of strong-coupling expansion.\cite{Pairault1,Pairault2,Brune1} Another drawback of the present approach is that the partition function can only be obtained indirectly by using the coupling constant integral method. The calculation of free energy is important for studying thermodynamical properties and constructing the conserving approximations. The present EOM-based expansion basically expands the excitation energies instead of the eigen energies. We have not yet found ways to construct the direct expansion of partition function.
  Finally, differing from the diagrammatic methods where a diagram in arbitrary order can be evaluated directly, in the present method, series can be generated only recursively and calculated order by order. 

In summary, we have presented an EOM-based method for doing series expansion of double time GFs. We developed both the bare expansion and the self-consistent expansion formula. Using this method, we carried out the second-order weak-coupling expansion as well as the strong-coupling expansion of the single particle GF for the single impurity Anderson model. For the weak-coupling expansion, Yamada's SE up to $U^2$ is obtained. For the strong-coupling expansion, we obtained results from three different expansion-resummation schemes: the bare expansion with SE resummation, bare expansion with CF resummation, and the self-consistent expansion with CF resummation. The latter overcomes both the causality problem and the zero-temperature divergence problem. We found that although they agree with NRG well in the large-$U$ and $T \gg \Gamma$ regimes, quantitative accuracy is not achieved at low temperature. Some features of this new approach and possible extensions are discussed.

\section{Acknowledgments}
This work was supported by the 973 Program of China (2012CB921704), NSFC Grant No. (11374362), Fundamental Research Funds for the Central Universities, and the Research Funds of Renmin University of China 15XNLQ03.

\appendix{}

\section{Calculation of $G_{1}(n_{\bar{\sigma}}d_{\sigma}|d_{\sigma}^{\dagger})_{\omega}$  in weak-coupling expansion }

In this appendix, we calculate $G_{1}(n_{\bar{\sigma}}d_{\sigma}|d_{\sigma}^{\dagger})_{\omega}$ using the bare EOM expansion. Straightforward calculation with the right-side EOM gives
\begin{eqnarray}   \label{A1}
&&  G_{1}(n_{\bar{\sigma}}d_{\sigma}|d_{\sigma}^{\dagger})_{\omega} =
G_{0}(d_{\sigma}| d_{\sigma}^{\dagger} )_{\omega} \times  \nonumber \\
&&  \left[\langle n_{\bar{\sigma}} \rangle_{1} + U G_{0}( n_{\bar{\sigma}} d_{\sigma}|  n_{\bar{\sigma}}d_{\sigma}^{\dagger} )_{\omega} - \alpha_{\sigma}G_{0}(  n_{\bar{\sigma}} d_{\sigma}| d_{\sigma}^{\dagger} )_{\omega} \right]. \nonumber \\
&&
\end{eqnarray}
This equation involves a zeroth-order two-particle GF $G_{0}( n_{\bar{\sigma}} d_{\sigma}|d_{\sigma}^{\dagger} )_{\omega}$ and a three-particle GF $G_{0}( n_{\bar{\sigma}} d_{\sigma}|  n_{\bar{\sigma}}d_{\sigma}^{\dagger} )_{\omega}$. $G_{0}( n_{\bar{\sigma}} d_{\sigma}|d_{\sigma}^{\dagger} )_{\omega}$ can be solved easily by its right-hand side EOM as $G_{0}(n_{\bar{\sigma}}d_{\sigma}|d_{\sigma}^{\dagger})_{\omega} = \langle n_{\bar{\sigma}} \rangle_{0} G_{0}(d_{\sigma}|d_{\sigma}^{\dagger})_{\omega}$. Direct EOM for the three-particle GF $G_{0}( n_{\bar{\sigma}} d_{\sigma}|  n_{\bar{\sigma}}d_{\sigma}^{\dagger} )_{\omega}$ will lead to new three-particle GFs and the closure of the hierarchy is slow. So we first diagonalize the unperturbed Hamiltonian $H_0$ (Eq.(\ref{eq:28}) in the main text) in the single-particle space and obtain
\begin{equation}  \label{A2}
    H_{0} = \sum_{s \sigma} \varepsilon_{s \sigma} a_{s\sigma}^{\dagger} a_{s \sigma}.
\end{equation}
Here $s$ is the single particle orbital index. We assume $d_{\sigma} = \sum_{s} h_{s \sigma} a_{s \sigma}$ with $\sum_{s} |h_{s \sigma}|^2 = 1$. Using the quasi-particle GF of $H_{0}$
\begin{equation}  \label{A3}
 G_{0}(a_{s \sigma}|a_{s' \sigma'}^{\dagger})_{\omega} = \frac{\delta_{s s'} \delta_{\sigma \sigma'} }{\omega - \varepsilon_{s\sigma} },
\end{equation} 
we express $G_{0}(d_{\sigma}|d_{\sigma'}^{\dagger})_{\omega}$ as 
\begin{eqnarray}  \label{A4}
G_{0}(d_{\sigma}|d_{\sigma'}^{\dagger})_{\omega} &=& \sum_{s s'} h_{s\sigma} h_{s' \sigma'}^{*}  G(a_{s\sigma}|a_{s' \sigma'}^{\dagger})_{\omega}   \nonumber \\
&=& \delta_{\sigma \sigma'} \sum_{s} \frac{|h_{s\sigma}|^2}{\omega - \varepsilon_{s\sigma}}.
\end{eqnarray}
The free LDOS is obtained as $\rho_{0 \sigma}(\epsilon) = -1/\pi {\text Im} G_{0}(d_{\sigma}|d_{\sigma}^{\dagger})_{\epsilon + i \eta} = \sum_{s} |h_{s\sigma}|^2 \delta(\epsilon - \varepsilon_{s\sigma})$.

Similarly, the two-particle GF is expressed in terms of the quasi-particle GFS as
\begin{eqnarray}  \label{A5}
 && G_{0}(n_{\bar{\sigma}}d_{\sigma}|n_{\bar{\sigma}}d_{\sigma}^{\dagger})_{\omega}  \nonumber \\
 & = & \displaystyle\sum_{suv} \displaystyle\sum_{s'u'v'} A_{suv, s'u'v'}  G_{0}( a_{s \bar{\sigma}}^{\dagger}  a_{u \bar{\sigma}} a_{v \sigma} | a_{s' \bar{\sigma}}^{\dagger}  a_{u' \bar{\sigma}} a_{v' \sigma}^{\dagger})_{\omega}, \nonumber \\
&& 
\end{eqnarray}
where 
$A_{suv, s'u'v'} = h_{s \bar{\sigma}}^{*} h_{u \bar{\sigma}} h_{v \sigma}  h_{s' \bar{\sigma}}^{*} h_{u' \bar{\sigma}} h_{v' \sigma}^{*}$.
The EOM for $G_{0}( a_{s \bar{\sigma}}^{\dagger}  a_{u \bar{\sigma}} a_{v \sigma} | a_{s' \bar{\sigma}}^{\dagger}  a_{u' \bar{\sigma}} a_{v' \sigma}^{\dagger})_{\omega}$ gives
\begin{eqnarray}  \label{A6}
&&   G_{0}( a_{s \bar{\sigma}}^{\dagger}  a_{u \bar{\sigma}} a_{v \sigma} | a_{s' \bar{\sigma}}^{\dagger}  a_{u' \bar{\sigma}} a_{v' \sigma}^{\dagger})_{\omega}  \nonumber \\
 & =& \frac{\langle \{a_{s \bar{\sigma}}^{\dagger}  a_{u \bar{\sigma}} a_{v \sigma}, a_{s' \bar{\sigma}}^{\dagger}  a_{u' \bar{\sigma}} a_{v' \sigma}^{\dagger}  \}  \rangle_{0}}{\omega + \varepsilon_{s\bar{\sigma}} - \varepsilon_{u \bar{\sigma}} -\varepsilon_{v\sigma} }.
\end{eqnarray}
The nominator of this GF is easily calculated as $\delta_{vv'}\delta_{us'}\delta_{su'} n_{v \sigma}(n_{u \bar{\sigma}} - n_{s \bar{\sigma}}) + \delta_{vv'}\delta_{us'}\delta_{su'} n_{s\bar{\sigma}}(1- n_{u \bar{\sigma}}) + \delta_{vv'}\delta_{us}\delta_{u's'} n_{s\bar{\sigma}} n_{s' \bar{\sigma}}$. Here
$n_{s \sigma} = \langle a_{s\sigma}^{\dagger}a_{s\sigma} \rangle_{0} = 1/(e^{\beta \varepsilon_{s \sigma}} + 1)$.
Putting it into Eq.(\ref{A6}) and (\ref{A5}), one obtains
\begin{eqnarray}  \label{A7}
&&   G_{0}( n_{\bar{\sigma}}  d_{\sigma} | n_{\bar{\sigma}} d_{ \sigma}^{\dagger})_{\omega}  \nonumber \\
 & =& \displaystyle\sum_{suv} \frac{ |h_{s\bar{\sigma}}|^2  |h_{u\bar{\sigma}}|^2 |h_{v \sigma}|^2 }{\omega + \varepsilon_{s\bar{\sigma}} - \varepsilon_{u \bar{\sigma}} -\varepsilon_{v\sigma} } \left[n_{v \sigma} \left( n_{u \bar{\sigma}} - n_{s \bar{\sigma}} \right) + n_{s \bar{\sigma}} \left(1 -n_{u \bar{\sigma}} \right) \right]  \nonumber \\
 && + \langle n_{\bar{\sigma}} \rangle_{0}^2 G_{0}(d_{\sigma}|d_{\sigma}^{\dagger})_{\omega}. 
\end{eqnarray}
In terms of the free LDOS, this equation is written as 
\begin{eqnarray}   \label{A8}
&&  G_{0}(n_{\bar{\sigma}}d_{\sigma}|n_{\bar{\sigma}}d_{\sigma}^{\dagger})_{\omega} = \langle n_{\bar{\sigma}}\rangle_{0}^2 G_{0}(d_{\sigma}|d_{\sigma}^{\dagger})_{\omega} \nonumber \\
&& + \displaystyle\iiint_{-\infty}^{\infty}  \frac{ \rho_{0 \bar{\sigma}}(\epsilon_1) \rho_{0 \bar{\sigma}}(\epsilon_2) \rho_{0 \sigma}(\epsilon_3) }{\omega + \epsilon_1 - \epsilon_2 -\epsilon_3} F(\epsilon_1, \epsilon_2, \epsilon_3) d\epsilon_1 d\epsilon_2 d\epsilon_3, \nonumber \\
&& 
\end{eqnarray}
with $F(\epsilon_1, \epsilon_2, \epsilon_3) = n_{\epsilon_3} \left( n_{\epsilon_2} - n_{\epsilon_1} \right) + n_{\epsilon_1} \left(1 -n_{\epsilon_2} \right)$. 
Finally, putting Eq.(\ref{A8}) into Eq.(\ref{A1}) and using the function $K_{\sigma}(\omega)$ defined in Eq.(\ref{eq:38}), we get
\begin{eqnarray}    \label{A9}
&&  G_{1}(n_{\bar{\sigma}}d_{\sigma}|d_{\sigma}^{\dagger})_{\omega} \nonumber \\
&&= \left[ \langle n_{\bar{\sigma}} \rangle_{1} +  U K_{\sigma}(\omega) \right] G_{0}(d_{\sigma}| d_{\sigma}^{\dagger} )_{\omega}  \nonumber \\
&&+ \langle n_{\bar{\sigma}} \rangle_{0} \left[ U \langle n_{\bar{\sigma}} \rangle_{0} - \alpha_{\sigma}\right] G_{0}^{2}(d_{\sigma}| d_{\sigma}^{\dagger} )_{\omega}. \nonumber \\
\end{eqnarray}
This completes the calculation of $G_{1}(n_{\bar{\sigma}}d_{\sigma}| d_{\sigma}^{\dagger})_{\omega}$.

\end{document}